\documentclass[%
reprint,
superscriptaddress,
preprintnumbers,
nobibnotes,
nofootinbib,
 amsmath,amssymb, 
 aps, prd,
 longbibliography,
]{revtex4-1}

\usepackage{cancel}
\usepackage{accents}
\usepackage{mciteplus,slashed}
\usepackage{amssymb,cancel,amsmath}
\usepackage{dcolumn}
\usepackage{bm}
\usepackage{soul}
\usepackage[caption=false]{subfig}
\usepackage{appendix}
\usepackage{physics}
\usepackage{booktabs}
\usepackage{comment}
\usepackage[T1]{fontenc}	
\usepackage{csvsimple}
\usepackage[bookmarksnumbered=true]{hyperref} 
\hypersetup{
    colorlinks = true,
    linkcolor = blue,
    anchorcolor = blue,
    citecolor = blue,
    filecolor = blue,
    urlcolor = blue
    }
\usepackage[section]{placeins}
\usepackage[capitalise]{cleveref}
\usepackage{booktabs}
\usepackage{graphicx}
\usepackage{mathrsfs}
\graphicspath{{Figures/}}
\AtBeginDocument{
\heavyrulewidth=.08em
\lightrulewidth=.05em
\cmidrulewidth=.03em
\belowrulesep=.65ex
\belowbottomsep=0pt
\aboverulesep=.4ex
\abovetopsep=0pt
\cmidrulesep=\doublerulesep
\cmidrulekern=.5em
\defaultaddspace=.5em
}
\usepackage[dvipsnames]{xcolor}
\usepackage[normalem]{ulem}
\usepackage{fontawesome} 
\usepackage[force]{feynmp-auto}
\usepackage{bbm}
\usepackage{MnSymbol}
\usepackage{comment}

\def\iu{\mathrm{i}}
\def\e{\mathrm{e}}

\definecolor{forestgreen}{HTML}{228B22}

\begin{document}

\title{Explicitly on-shell currents in relativistic mean field models}

\author{Alexis Nikolakopoulos}
\affiliation{Department of Physics, University of Washington, Seattle, Washington 98195, USA}

\author{Ryan Plestid}
\affiliation{CERN, 1 Esplanade des Particules, CH-1211 Geneva 23, Switzerland}

\date{\today}

\preprint{CERN-TH-2026-182}
\preprint{NT@UW-26-18}

\begin{abstract}
    Relativistic mean field models are a useful tool for modeling semi-leptonic scattering and photo production on nuclei. When using free-nucleon currents, it is often claimed that there exist so-called ``off-shell ambiguities''. Here we show that when the current is defined in terms of free-nucleon creation and annihilation operators, all  ambiguities related to on-shell vs. off-shell Dirac algebra disappear. Genuine ambiguities persist because the current itself depends on the mean field responsible for nuclear binding; these would be fixed if a consistent background-field dependent current were used.  As applications, we consider elastic scattering from a nucleus, and (very large) ambiguities that were previous reported in the literature in the context of coherent pion photoproduction. We explain how these ambiguities are removed by the procedure introduced herein. 
\end{abstract}

 \maketitle 

\section{Introduction}
Relativistic mean field models that treat nuclei as a collection of nucleons bound in a background potential (composed of scalar and vector fields) are phenomenologically successful and computationally inexpensive~\cite{RINGRMF, walecka04, Serot:1997xg, BenderReinhard:RevModPhys, Reinhard_1989}. By formulating the problem in terms of a relativistic field theory (i.e., at the level of the Dirac equation) relativistic corrections are automatically incorporated when modeling high energy processes such as quasi-elastic scattering or high energy photo-production of mesons. 

Practitioners of mean field models often encounter so-called ``off-shell ambiguities''. These arise because the kinematics in scattering on a nucleus differ from scattering on a free nucleon. The typical prescription in a mean field model is to solve the Dirac equation and obtain a set of orbital wavefunctions $\phi(\vb{x})$ where $\phi$ is a four-component spinor. Next, one takes a hadronic matrix element for free nucleon states, 
\begin{equation}
    \label{current-illustration}
    \mel{N(\vb{p}',s')}{\hat{J}_\mu}{N(\vb{p},s)} = \bar{u}(\vb{p}',s') \qty[\sum_i F_i(Q^2) \Gamma^\mu_i ] u(\vb{p},s)~,
\end{equation}
where $F_i(Q^2)$ are form factors depending on $Q^2=(E_{p'}-E_{p})^2 - (\vb{p}'-\vb{p})^2$ and $\Gamma_i$ are an appropriate basis of Dirac bilinears.  Finally, for matrix elements between states (either bound or continuum) in the presence of the mean-field model's background fields, one uses the prescription
\begin{equation}
    \label{prescription}
    \mel{\beta}{\hat{J}_\mu(\vb{q})}{\alpha} \rightarrow \int \frac{\dd^3 \vb{p}}{(2\pi)^3} \overline{\phi}_\beta(\vb{p}+\vb{q}) \qty[\sum_i F_i(Q^2) \Gamma^\mu_i ] \phi_\alpha(\vb{p})~. 
\end{equation}
Such an identity is known to hold for local currents of elementary fields in (e.g., electrons bound in an external Coulomb potential); we comment on this example in detail below. 

While this procedure seems sensible, one encounters alarming ambiguities. The choice of bilinears in \cref{current-illustration} is not unique, for example one can use the Gordon identity to trade $\gamma_\mu$ for a linear combination of $(p+p')_\mu$ and $\sigma_{\mu\nu}(p-p')^\nu$. Yet, one readily finds that results are basis dependent when inserting these different choices of bilinears between bound or continuum eigenstates using \cref{prescription}.  This is an example of an ``off-shell ambiguity'', as discussed in the literature~\cite{Caballero:1997gc, DEFOREST1983232, CABALLERO1993709}.

These ambiguities are not an entirely academic issue. For example, in the context of coherent pion production they were found to lead to $\sim 500\%$ uncertainties in the predicted cross section \cite{Abu-Raddad:1998ams}. This clearly undermines any hope of predictivity for a relativistic mean field model. Addressing this issue is particularly relevant in view of the recent applications of the theory to the analysis of neutrino experiments and the adoption of this approach in event generators~\cite{Nikolakopoulos:2022tut, Nikolakopoulos:2022qkq, Gonzalez-Jimenez19, McKean:2025khb, Nikolakopoulos:2024mjj, McKean:2026odw,Butkevich:2025fsf, Dolan:2026dfu, Franco-Patino:2026rwy, Franco-Munoz:2026hxi}.

In principle these ambiguities may stem from three logically distinct sources. For concreteness consider an electromagnetic current, $\hat{J}_\mu$, from the three-point vertex with a photon:
\begin{enumerate} 
    \item The current operator, $\hat{J}_\mu$, will in general depend on background fields used to define the bound or continuum states $\ket{\alpha}$ and $\ket{\beta}$. With on-shell information from free nucleon scattering one only has access to the current in the absence of background fields.
    \item There is an ambiguity in the operator level definition of the current.  From on-shell amplitudes alone one cannot discern if one should take $Q^\mu = (E_{\vb{p}'}-E_{\vb{p}}, \vb{q})$ or $q^\mu=(\omega, \vb{q})$ with $\omega$ the incoming photon energy in a bound state and $E_{\vb{p}}=\sqrt{\vb{p}^2+M^2}$. 
    \item Spinor algebra which holds on-shell allows for different representations of the same on-shell amplitudes. These different representations give different answers when evaluated using \cref{prescription}.    
\end{enumerate}
The first issue is a genuine limitation of using free-nucleon matrix elements in a mean-field models. The second point is a genuine ambiguity but can be ``fixed'' if one has an auxiliary principle\footnote{For example an effective Lagrangian description often fixes the operator structure in the low-$\vb{q}^2$ limit.} that fixes the operator definition of the current. The third issue, as we will now argue, is spurious.

In what follows, we show that for currents defined in terms of free-nucleon creation and annihilation operators, no off-shell ambiguities of the third-type exist. The prescription \cref{prescription} is correct for elementary currents of the form $\hat{J} = \bar{\psi} \Gamma \psi$, with $\psi$ the nucleon field operator and $\Gamma$ independent of momenta, but is incorrect for realistic nucleon currents. If the current operator is defined at the operator-level, including background field dependence that ensures that Ward identities are satisfied, then no ambiguities exist. 

The rest of the paper is organized as follows. In \cref{sec:MF-models} the notion of a relativistic mean-field model is defined. Next, in \cref{sec:currents} we define current operators in terms of free-nucleon creation and annihilation operators (this does not account for the dependence of currents on the mean-field potential). We  explain how they act in bound states, and how off-shell ambiguities disappear when a careful operator level treatment is followed. We apply these ideas to phenomenology in \cref{sec:phenomenology} focusing on coherent elastic scattering, and coherent pion production. Finally in \cref{sec:conclusions} we summarize our findings and comment on potential future directions. 

\section{Mean-field models \label{sec:MF-models} }
In the absence of external currents, the mean field model is a quadratic theory with time-independent classical vector and scalar fields, 
\begin{equation}
    \mathcal{L} = \bar{\psi}\qty(-\iu\slashed{\partial} + \gamma_0 V(\vb{x}) + \qty[m_N + S(\vb{x})]) \psi ~.  
\end{equation}
Variation of the action gives the equations of motion for the field operator $\hat{\psi}(\vb{x},t)$ in the Heisenberg picture, 
\begin{equation}
    \qty(-\iu\slashed{\partial} + \gamma_0 V(\vb{x}) + \qty[m_N + S(\vb{x})])\hat{\psi}(t,\vb{x}) =0 ~.
\end{equation}
For any configuration of classical fields, there exists a minimum energy state (the vacuum) $\ket{\Omega[S,V]}$ which is a functional of the background fields. We may renormalize our Hamiltonian (for any fixed configuration of $V$ and $S$) such that $\hat{H}\ket{\Omega}=0$. In the absence of scalar or vector fields we recover the free-vacuum $\ket{0}=\ket{\Omega[0,0]}$. 

We may turn the Heisenberg equations of motion for $\hat{\psi}$ into a $c$-number differential equation by taking a matrix element between the vacuum and a one-particle state labeled by $\alpha$, 
\begin{equation}
    \phi_\alpha(\vb{x},t)=\mel{\alpha}{\hat{\psi}(\vb{x},t)}{\Omega}= \e^{\iu E_\alpha t} \mel{\alpha}{\hat{\psi}(\vb{x},0)}{\Omega}~,
\end{equation}
where we have translated the field to $t=0$. We then obtain, 
\begin{equation}
    \label{Dirac-equation-phi-alpha}
    \qty(E_\alpha \gamma_0  + \gamma_0 V(\vb{x}) + \iu \vb*{\gamma}\cdot \vb*{\nabla} + \qty[m_N + S(\vb{x})])\phi_\alpha(\vb{x})=0~. 
\end{equation}
A similar equation is obtained for single particle states containing a valence anti-nucleon, 
\begin{equation}
    \chi_{\mathfrak{a}}(\vb{x},t)\mel{\Omega}{\hat{\psi}(\vb{x},t)}{\mathfrak{a}}= \e^{-\iu E_{\mathfrak{a}} t} \mel{\Omega}{\hat{\psi}(\vb{x},0)}{\mathfrak{a}}~.
\end{equation}
Notice that the energy $E_{\mathfrak{a}}$ is positive, but enters with a negative sign since we act on the anti-nucleon state $ \ket{\mathfrak{a}}$. We then arrive at the equations for negative eigenvalue solutions, 
\begin{equation}
    \qty(-E_{\mathfrak{a}} \gamma_0  + \gamma_0 V(\vb{x}) + \iu \vb*{\gamma}\cdot \vb*{\nabla} + \qty[m_N + S(\vb{x})])\chi_{\mathfrak{a}}(\vb{x})=0~. 
\end{equation}

At $t=0$ the Schr\"odinger and Heisenberg pictures agree, and the field operator, $\hat{\psi}(\vb{x},0)$, can be expanded in any convenient basis. For example given the standard free-nucleon creation and annihilation operators (defined such that $\hat{b}_{\vb{p},\sigma}\ket{0}=0$ and $\hat{d}_{\vb{p},\sigma}\ket{0}=0$), we have 
\begin{equation}
    \begin{split}
    \hat{\psi}(\vb{x},0)=\sum_s \int &\frac{\dd^3p}{(2\pi)^3 \sqrt{2 E_p}} \\
    &\qty[\hat{b}_{\vb{p},s} u(\vb{p},s)\e^{\iu \vb{p}\cdot\vb{x}} + \hat{d}^\dagger_{\vb{p},s} v(\vb{p},s)\e^{-\iu \vb{p}\cdot\vb{x}} ]~.
    \end{split}
\end{equation}
When working with non-zero background fields it is useful to work in terms of the operators that diagonalize the Hamiltonian. These operators satisfy $\hat{c}_\alpha \ket{\Omega} =0$ and $\hat{\mathfrak{c}}_{\mathfrak{a}}\ket{\Omega}=0$, and lead to a mode-decomposition of the form 
\begin{equation}
   \hat{\psi}(\vb{x})=  \sum_\alpha  \phi_\alpha(\vb{x}) \hat{c}_\alpha + \sum_{\mathfrak{a}}   \chi_{\mathfrak{a}}(\vb{x}) \hat{\mathfrak{c}}^\dagger_{\mathfrak{a}}~,
\end{equation}
where the sum is schematic and includes both bound- and continuum states. 

Using these two representations, we can relate the two sets of creation and annihilation operators to one another via a Bogoliubov transformation. For example, using $u^\dagger(-\vb{p},\sigma ) v(\vb{p},s) = v^\dagger(-\vb{p},\sigma)u(\vb{p},s)=0$ and $u^\dagger(\vb{p},\sigma)u(\vb{p},s)= u^\dagger(\vb{p},\sigma)u(\vb{p},s) = 2 E_{p} \delta_{\sigma, s}$ we have \cite{Hoyer:2021adf}, 
\begin{align}
    \hat{b}_{\vb{p},s} &=  \int \dd^3x ~\frac{1}{\sqrt{2 E_p}}\e^{-\iu \vb{p}\cdot \vb{x}} u^\dagger(\vb{p},s) \hat{\psi}(\vb{x}) \\
    &= \sum_\alpha u^\dagger(\vb{p},s)\tilde{\phi}_\alpha(+\vb{p})\hat{c}_\alpha + \sum_{\mathfrak{a}} u^\dagger(\vb{p},s)\tilde{\chi}_{\mathfrak{a}}(-\vb{p})\hat{\mathfrak{c}}_{\mathfrak{a}}^\dagger ~,\nonumber \\
    \hat{d}_{\vb{p},s}^\dagger &=  \int \dd^3x ~\frac{1}{\sqrt{2 E_p}}\e^{+\iu \vb{p}\cdot \vb{x}} v^\dagger(\vb{p},s) \hat{\psi}(\vb{x}) \\
    &= \sum_\alpha v^\dagger(\vb{p},s)\tilde{\phi}_\alpha(-\vb{p})\hat{c}_\alpha + \sum_{\mathfrak{a}} v^\dagger(\vb{p},s)\tilde{\chi}_{\mathfrak{a}}(+\vb{p}) \hat{\mathfrak{c}}_{\mathfrak{a}}^\dagger~, \nonumber 
\end{align}
where the tilde denotes a Fourier transform.

\pagebreak

Similarly, we can obtain the inverse relations. For example the positive and negative energy creation and annihilation operators can be written in terms of the free solutions as, 
\begin{align}
    \hat{c}_\alpha^\dagger &= \int \dd^3 x ~   \hat{\psi}^\dagger (\vb{x})\phi_\alpha(\vb{x}) \nonumber \\
    &= \sum_s \int \frac{\dd^3 p}{(2\pi)^3} \frac{1}{\sqrt{2 E_p}} \\
    &\hspace{0.1\linewidth}\times \qty[u^\dagger(\vb{p},s) \tilde{\phi}_\alpha(\vb{p}) \hat{b}_{\vb{p},s}^\dagger +  v^\dagger(\vb{p},s)
    \tilde{\phi}_\alpha(-\vb{p}) \hat{d}_{\vb{p},s} ]~,  \nonumber 
\end{align}
and
\begin{align}
    \hat{\mathfrak{c}}_a &= \int \dd^3 x   ~\hat{\psi}^\dagger(\vb{x})\chi_a(\vb{x})~ \nonumber \\
    &= \sum_s \int \frac{\dd^3 p}{(2\pi)^3} \frac{1}{\sqrt{2 E_p}} \\
    &\hspace{0.1\linewidth}\times \qty[ u^\dagger(\vb{p},s) \tilde{\chi}_a(\vb{p})\hat{b}_{\vb{p},s}^\dagger + v^\dagger(\vb{p},s)
    \tilde{\chi}_a(-\vb{p})  \hat{d}_{\vb{p},s} ]~.\nonumber
\end{align}
From these relations it is manifest that $\ket{\Omega} \neq \ket{0}$, however one can construct $\ket{\Omega}$ in terms of free-particle states using an operator which produces correlated nucleon anti-nucleon pairs. Up to a normalization factor $N_0$ we have,  
\begin{widetext}
\begin{align}
    \ket{\Omega} &= N_0 \exp\qty[ - \sum_{s,\sigma}\int \frac{\dd^3 p}{(2\pi)^3} \frac{1}{\sqrt{2E_p}}  \int \frac{\dd^3 q}{(2\pi)^3} \frac{1}{\sqrt{2E_q} }\hat{b}^\dagger_{\vb p,s} K_{s,\sigma}(\vb{p},\vb{q}) \hat{d}^\dagger_{\vb{q},\sigma}] \ket{0}~,\\
    K_{s,\sigma}(\vb{p},\vb{q})&=  \frac{1}{\sqrt{2 E_p}} \frac{1}{\sqrt{2 E_q}} 
    \sum_{\alpha} \qty[\tilde{\phi}_\alpha^\dagger(-\vb{p}) v(\vb{p},s)] \qty[u^\dagger(\vb{q},\sigma)\tilde{\phi}_\alpha(\vb{q})]~. \label{eq:interacting_vacuum}
\end{align}
\end{widetext}
Expanded out one will find the vacuum state, $\ket{\Omega[V,S]}$, contains terms with no free-nucleons $\ket{0}$, and pairs of nucleons and anti-nucleons e.g., $\ket{N\bar{N}}$ \cite{Hoyer:2021adf}. A similar phenomenon appears when studying Compton scattering in an explicitly time-ordered formalism (see \cref{app:compton-example}).

\vspace{-12pt}
\section{Current operators \label{sec:currents} }
Next let us consider the definition of a current $\hat{J}_\mu$. In QCD, the current assumes a simple form in terms of quark bilinears $\sum_q \bar{q}\Gamma q$, and has many non-trivial matrix elements. For example $\mel{N}{\bar{q}\Gamma q}{N}$ and $\mel{N \pi}{\bar{q}\Gamma q}{N}$ are mediated by the same microphysical degrees of freedom. 

In our mean-field model we have a theory of non-interacting nucleons. Therefore, to obtain realistic current matrix elements, we will need to supply form factors ``by hand'' as well as any interactions with e.g., pions.  This can be straightforwardly achieved by expanding the current as 
\begin{equation}
    \hat{J}_\mu= \hat{J}_\mu^{(1)} + \hat{J}_\mu^{(1)\pi} + \ldots~, 
\end{equation}
where higher order terms may include two-body currents and other final state particles. 

The one-body current $\hat{J}_\mu^{(1)}$ may be {\it defined} so as to reproduce all relevant nucleon matrix elements, 
\begin{equation}    
    \label{current-def}
    \begin{split}
    \hat{J}_\mu^{(1)} = &\sum_{s,s'}\int \frac{\dd^3 p}{(2\pi)^3}\frac{1}{\sqrt{2 E_p}}\frac{\dd^3 p'}{(2\pi)^3}\frac{1}{\sqrt{2 E_{p'}}} \\
    \bigg[ &\phantom{+~} 
    \mel{N(\vb{p}',s')}{\hat{J}_\mu}{N(\vb{p},s)} b^\dagger_{\vb{p}',s'} b_{\vb{p},s} \\
    &+  \mel{\bar{N}(\vb{p},s)}{\hat{J}_\mu}{\bar{N}(\vb{p}',s')}  d_{\vb{p},s}d^\dagger_{\vb{p},s}\\
    &+ \mel{\bar{N}(\vb{p}',s')N(\vb{p},s)}{\hat{J}_\mu}{0} b_{\vb{p},s}^\dagger
    d_{\vb{p},s}^\dagger \\
    &+\mel{0}{\hat{J}_\mu}{\bar{N}(\vb{p}',s')N(\vb{p},s)} d_{\vb{p}',s'} b_{\vb{p},s}~~\bigg]~.
    \end{split}
\end{equation}
Notice that we have chosen not to normal order $d d^\dagger$ to match what one would find by expanding $\bar{\psi}\Gamma\psi$. It is important to emphasize that this definition is {\it a model} for the current. It has not been derived microscopically, and we have imposed (by hand) the constraint that the current contains only nucleon bilinears and does not depend on the background fields $S(\vb{x})$ and $V(\vb{x})$. As we discuss below, this leads to issues related to current conservation.

The price we pay for realistic matrix elements is a non-local current with both space- and time-like form factors
\begin{align}
    \mel{N(\vb{p}',s')}{\hat{J}_\mu}{N(\vb{p},s)}             &=  
                                    \bar{u}(\vb{p}',s') \Gamma_\mu^{(a)}(\vb{p}',\vb{p}) u(\vb{p},s)~,\nonumber \\
    \mel{\bar{N}(\vb{p},s)}{\hat{J}_\mu}{\bar{N}(\vb{p}',s')} &= 
                                    (-1)\bar{v}(\vb{p}',s') \Gamma_\mu^{(b)}(\vb{p}',\vb{p}) v(\vb{p},s) ~,\nonumber \\
    \mel{\bar{N}(\vb{p}',s')N(\vb{p},s)}{\hat{J}_\mu}{0}      &=  
                                    \bar{u}(\vb{p}',s') \Gamma_\mu^{(c)}(\vb{p}',\vb{p}) v(\vb{p},s) ~, \nonumber\\
    \mel{0}{\hat{J}_\mu}{\bar{N}(\vb{p}',s')N(\vb{p},s)}      &=  
                                    \bar{v}(\vb{p}',s') \Gamma_\mu^{(d)}(\vb{p}',\vb{p}) u(\vb{p},s)~.
                                    \label{eq:def_bilinears}
\end{align}
We have included an explicit factor of $(-1)$ in the second equation which arises from reordering fields prior to Wick contractions with the external states. 

The reason why time-like channels are relevant for the current is because the vacuum $\ket{\Omega}$ contains free-particle Fock states such as $\ket{N\bar{N}}$. By extension, single particle states $\ket{\alpha}=\hat{c}_\alpha^\dagger \ket{\Omega[V,S]}$ contain higher multiplicity free-particle states such as $\ket{N N\bar{N}}$. It follows that matrix elements of the form $\mel{0}{\hat{J}_\mu}{N\bar{N}}$ can appear in calculations of $\mel{\beta}{\hat{J}_\mu}{\alpha}$ (see also the discussion in Ref.~\cite{Brodsky:1983ta}). 

\vspace{-12pt}

\subsection{Elementary currents}
\label{sec:elementary_my_dear}
A special case of the preceding discussion are local currents defined without form factors. It is well established in theories of elementary fermions (e.g., electrons bound in a background Coulomb field) that the prescription of \cref{prescription} is exact. We should be able to recover this statement using our definition of the current. We will see this is indeed the case, and that it relies on all four channels (two time-like, and two space-like). 

In what follows, we will flip three-momenta, and it is convenient to introduce
\begin{equation}
    p_\mu^{(\prime)}=(E^{(\prime)},+\vb{p}^{(\prime)})  ~~,~~ \tilde{p}_\mu^{(\prime)} = (E^{(\prime)},-\vb{p}^{(\prime)}) ~, 
\end{equation}
with $E^{(\prime)} = \sqrt{p^{(\prime)2} +M^2}$. Discarding disconnected graphs, we have 
\begin{equation}
         \mel{\beta}{\bar{\psi}\Gamma \psi}{\alpha} =  \sum_{s,s'}\int \frac{\dd^3 p}{(2\pi)^3}\frac{1}{2 E_p}\frac{\dd^3 p'}{(2\pi)^3}\frac{1}{2 E_{p'}} \qty[ \ldots ]~,
\end{equation}
with the term in brackets being given by 
\begin{equation}
    \nonumber
    \begin{split}
     \bigg[&
    \tilde{\phi}^\dagger_{\beta} (\vb{p}')u(\vb{p}',s') \bar{u}(\vb{p}',s') \Gamma u(\vb{p},s) u^\dagger(\vb{p},s) \tilde{\phi}_\alpha(\vb{p})\\
    &+ \tilde{\phi}^\dagger_{\beta} (-\vb{p}')v(\vb{p}',s') \bar{v}(\vb{p}',s') \Gamma v(\vb{p},s) v^\dagger(\vb{p},s) \tilde{\phi}_\alpha(-\vb{p})\\
    &+ \tilde{\phi}^\dagger_{\beta} (-\vb{p}')v(\vb{p}',s') \bar{v}(\vb{p}',s') \Gamma u(\vb{p},s) u^\dagger(\vb{p},s) \tilde{\phi}_\alpha(\vb{p})\\
    &+ \tilde{\phi}^\dagger_{\beta} (\vb{p}')u(\vb{p}',s') \bar{u}(\vb{p}',s') \Gamma v(\vb{p},s) v^\dagger(\vb{p},s) \tilde{\phi}_\alpha(-\vb{p})\bigg]~.
    \end{split}   
\end{equation}
By re-labelling momenta $\vb{p}^{(\prime)} \rightarrow - \vb{p}^{(\prime)}$ in appropriate places, and performing spin sums, we find
\begin{equation}
    \begin{split}
     \mel{\beta}{\bar{\psi}\Gamma \psi}{\alpha} &=  \int \frac{\dd^3 p}{(2\pi)^3}\frac{1}{2 E_p}\frac{\dd^3 p'}{(2\pi)^3}\frac{1}{2 E_{p'}}\\
     &\bigg[
    \tilde{\phi}^\dagger_{\beta} (\vb{p}')(\slashed{p}'+m) \Gamma(\slashed{p} +m)\gamma_0 \phi_\alpha(\vb{p})\\
    &\hspace{0.025\linewidth}+ \tilde{\phi}^\dagger_{\beta} (\vb{p}')(\tilde{\slashed{p}}'-m)\Gamma (\tilde{\slashed{p}}-m)\gamma_0 \phi_\alpha(\vb{p})\\
    &\hspace{0.025\linewidth}+ \tilde{\phi}^\dagger_{\beta} (\vb{p}')(\tilde{\slashed{p}}'-m)\Gamma (\slashed{p}+m)\gamma_0 \phi_\alpha(\vb{p})\\
    &\hspace{0.025\linewidth}+ \tilde{\phi}^\dagger_{\beta} (\vb{p}') (\slashed{p}'+m) \Gamma (\tilde{\slashed{p}}-m) \gamma_0 \phi_\alpha(\vb{p})\bigg]~.
    \end{split}   
\end{equation}
Then using $(\tilde{\slashed{p}}-m)+(\slashed{p}+m)=2 E_p \gamma_0$ we arrive at, 
\begin{equation}
    \label{elementary-current}
    \mel{\beta}{\bar{\psi}\Gamma \psi}{\alpha} = \int \frac{\dd^3 p'}{(2\pi)^3} \int \frac{\dd^3 p}{(2\pi)^3} \tilde{\phi}_\beta^\dagger(\vb{p}') \gamma_0 \Gamma \tilde{\phi}_\alpha(\vb{p})~,
\end{equation}
as expected. This same relation can be obtained by performing operator algebra at the level of the current. 

In order to derive \cref{elementary-current}, we had to sum the time-like and space-like channels together. For the above derivation it was key that these both have the same Dirac bilinear, $\Gamma$, and that $\Gamma$ not depend on momentum; let us generalize further. Consider the interaction Hamiltonian of a nucleon field $\psi$ written as, 
\begin{equation}
\label{eq:Hint_elementary}
    H_{\rm int} = g_1 \bar{\psi}\gamma_\mu \psi A^\mu + g_2 \bar{\psi} \sigma_{\mu\nu} \psi F^{\mu\nu}~,
\end{equation}
where $F_{\mu\nu} = \partial_\mu A_\nu - \partial_\nu A_\mu$ with $A_\mu$ the photon field. Notice that because we have written the interaction in terms of $F^{\mu\nu}$, and $g_{1,2}$ and simply real numbers, there are no form factors. Furthermore, the relevant Dirac bilinears have no momentum dependence. 

If, however, we use the Gordon identity, 
\begin{equation}
    \bar{u}(p') \iu \sigma^{\mu\nu} q_\nu u(p) = \bar{u}(p')[2M \gamma^\mu - (p+p')^\mu] u(p)~, 
\end{equation}
then we find pieces proportional to $(p+p')_\mu$. These terms require additional care, and one must ``promote'' $\Gamma\rightarrow \Gamma(\vb{p}', \vb{p})$. In this more general scenario, one finds,
\begin{widetext}
\begin{equation}
    \begin{split}\label{main-result}
     \mel{\beta}{\hat{J}_\mu^{(1)}}{\alpha} =  \int &\frac{\dd^3 p}{(2\pi)^3}\frac{1}{2 E_p}\frac{\dd^3 p'}{(2\pi)^3}\frac{1}{2 E_{p'}}\tilde{\phi}^\dagger_{\beta}(\vb{p}')\bigg[
    (\slashed{p}'+m) \Gamma_\mu^{(a)}(\vb{p}',\vb{p})(\slashed{p} +m)\gamma_0 
    + (\tilde{\slashed{p}}'-m)\Gamma_\mu^{(b)}(-\vb{p}',-\vb{p})(\tilde{\slashed{p}}-m)\gamma_0 \\
    &\hspace{0.15\linewidth}+  (\tilde{\slashed{p}}'-m)\Gamma_\mu^{(c)}(-\vb{p}',\vb{p}) (\slashed{p}+m)\gamma_0 +  (\slashed{p}'+m) \Gamma_\mu^{(d)}(\vb{p}',-\vb{p}) (\tilde{\slashed{p}}-m) \gamma_0 \bigg]\tilde{\phi}_\alpha(\vb{p})~.
    \end{split}   
\end{equation}
\end{widetext}

\Cref{main-result} gives the same answer no matter which equivalent on-shell form of the current is used. There are projectors in place such that all on-shell algebra (e.g., the Gordon identity or other techniques that use the free Dirac equation) operates undisturbed and no ``off-shell ambiguities'' appear. If, however, one uses \cref{elementary-current} naively, one obtains {\it different} answers. 
Let us illustrate this by considering $\Gamma^{(a,b,c,d)} = \gamma^\mu$. The Gordon identity for each contribution yields
\begin{align}
\label{eq:Gdecompuu}
\overline{u}(p^\prime) \gamma^\mu u(p) &= \overline{u}(p^\prime)\left[ \frac{p^{\prime\mu} + p^\mu}{2M} + \iu\frac{\sigma^{\mu\nu}}{2M}\left(p^\prime - p \right)_\nu \right] u(p) ~, \\
\overline{v}(p^\prime) \gamma^\mu v(p) &= (-1) \overline{v}(p^\prime)\left[ \frac{p^{\prime\mu} + p^\mu}{2M} + \iu\frac{\sigma^{\mu\nu}}{2M}\left(p^\prime - p \right)_\nu \right] v(p) ~, \label{eq:Gdecompvv} \\
\overline{v}(p^\prime) \gamma^\mu u(p) &= (-1) \overline{v}(p^\prime)\left[ \frac{p^{\prime\mu} - p^\mu}{2M} + \iu\frac{\sigma^{\mu\nu}}{2M}\left(p^\prime + p \right)_\nu \right] u(p)~, \\
\overline{u}(p^\prime) \gamma^\mu v(p) &= \overline{v}(p^\prime)\left[ \frac{p^{\prime\mu} - p^\mu}{2M} + \iu\frac{\sigma^{\mu\nu}}{2M}\left(p^\prime + p \right)_\nu \right] v(p)~. 
\label{eq:Gdecompuv}
\end{align}
We can then rewrite the $(b)$-term in \cref{main-result} for example by using the bilinear 
\begin{equation}
    \Gamma^{(b)}(-\vb{p}^\prime,-\vb{p}) =  -\frac{\left(\tilde{p}^\prime + \tilde{p} \right)^\mu}{2M} -\frac{\iu}{2M}\sigma^{\mu\nu}\left(\tilde{p}^\prime - \tilde{p} \right)_\nu,
\end{equation}
instead of $\gamma^\mu$.
One can make (or not make) a similar substitution for any of the other contributions, the final result is the same.\!\footnote{One can explicitly verify that each the forms of \crefrange{eq:Gdecompuu}{eq:Gdecompuv} together with the appropriate sign changes of the momenta in \cref{main-result} are indeed equivalent to using $\gamma^\mu$. }
However, for the form of \cref{elementary-current} to be valid we need to choose a set of bilinears that satisfy
\begin{equation}
\label{eq:condition}
    \Gamma^{(a)}(\vb{p}^\prime,\vb{p}) = \Gamma^{(b)}(-\vb{p}^\prime, -\vb{p}) = \Gamma^{(c)}(-\vb{p}^\prime,\vb{p}) = \Gamma^{(b)}(\vb{p}^\prime, -\vb{p}).
\end{equation}
Clearly this is not the case for the right-hand sides of \crefrange{eq:Gdecompuu}{eq:Gdecompuv}.
By inspection of these equations, one sees what goes wrong when one erroneously uses \cref{elementary-current} with $\Gamma^\mu$ obtained from the Gordon decomposition given on the right-hand side of \cref{eq:Gdecompuu}.
The $\Gamma^{(b)}$ contribution changes sign. Or, put equivalently, a spurious contribution $\Gamma^{(b)} \rightarrow \Gamma^{(b)}-2\gamma^\mu$ is introduced in this term.
The $\Gamma^{(d)}$ term receives a spurious contribution $(\iu \sigma^{\mu\nu} p _\nu - p_\nu)/M$, and $\Gamma^{(c)} \rightarrow \Gamma^{(c)} -(\iu \sigma^{\mu\nu} p^\prime_\nu - p^{\prime,\mu})/M $.

\subsection{Currents with form factors}
\label{sec:Currents_w_FF}
When an irreducible momentum dependence is present, e.g. form factors or particle propagators, the results do not re-assemble into the naively expected form.  In this case \cref{main-result} needs to be employed.
Let us highlight some interesting features of this equation. The projectors for anti-particles involve a $(\tilde{\slashed{p}}-m)$ and the wavefunction evaluated at $+\vb{p}$ (equivalently we could write the answer in terms of $(\slashed{p}-m)$ and evaluate the wavefunction at $-\vb{p}$). The expected factor of $\gamma_0$ appears not beside $\tilde{\phi}^\dagger_\beta$ but instead beside $\tilde{\phi}_\alpha$. 
Based on this equation we may define positive/negative energy projections of the wavefunctions as
\begin{equation}
    \phi_{\alpha,+}(\vb{p}) = \frac{(\slashed{p}+M)}{2E} \gamma^0 \tilde{\phi}_\alpha(\vb{p}), \quad \phi_{\alpha,-}(\vb{p}) = \frac{(\slashed{\tilde{p}}-M)}{2E} \gamma^0 \tilde{\phi}_\alpha(\vb{p}) \nonumber
\end{equation}
Each of the contributions in \cref{main-result} is then of a familiar form e.g.  $\overline{\psi}_{\beta,+} \Gamma^{(a)} \psi_{\alpha,+}$.

The main complication that arises is that a consistent treatment of the $\Gamma^{(c,d)}$ is that these now involve time-like form factors.
It is worth noting that for binding energies small compared to the rest mass, there is a stark hierarchy  
\begin{equation}
    v^\dagger(-\vb{p})\phi_\alpha \ll u^\dagger(\vb{p}) \phi_\alpha~. 
\end{equation}
In fact, the left-hand side of this inequality may be estimated as being $O\qty(\langle \vb{p}^2 \rangle^{3/2}/M^3)$ relative to the right-hand side, where $\langle p \rangle $ is typical momentum scale in the bound state \cite{Plestid:2024xzh}. 

This is a consequence of the fact that for weak binding, the positive energy solutions are {\it very} similar to a linear superposition of free-nucleon states. Neglecting all suppressed terms, the leading approximation is hence given by,
\begin{equation}
         \mel{\beta}{\hat{J}_\mu^{(1)}}{\alpha} \simeq \int \frac{\dd^3 p}{(2\pi)^3}\frac{\dd^3 p'}{(2\pi)^3}
     \overline{\phi}_{\beta,+}(\vb{p}') \Gamma_\mu^{(a)}(\vb{p}',\vb{p})\phi_{\alpha,+}(\vb{p})~.\nonumber
\end{equation}
This involves only the nucleon space-like form factors, and is manifestly independent of the basis chosen to describe the free-nucleon form factors.

The positive-energy four-component wavefunctions are of the form
\begin{equation}
    \phi_{\alpha,+}(\vb{p}) = u_+(\vb{p}) \chi_{\alpha,+}(\vb{p}),
\end{equation}
with $u_{+}$ a positive-energy spinor (projector) and $\chi_{\alpha,+}$ a two-component wavefunction. Explicit expressions for bound states are given in \cref{app:useful-formulae}.
This means that the $++$ contribution may be identified with the ``traditional'' nuclear physics treatment in which $\overline{u}(\vb{p}) \Gamma(\vb{p}^\prime, \vb{p}) u(\vb{p})$ is considered in order to define a set of single-nucleon operators. One typically performs an expansion in powers of $p/M$ and identifies terms proportional to nuclear multipole operators~\cite{walecka04, Waleckapaper}.
In matching \cref{main-result} to such a procedure, the $\Gamma^{(a)}$ gives the leading contribution, while a consistent treatment of the $\Gamma^{(b,c,d)}$ allows to include higher order relativistic corrections. 

Lastly, notice that \cref{main-result} differs from past treatments~\cite{Caballero:1997gc, Nikolakopoulos:2025cku} that start from \cref{prescription} and inserting projectors $(\slashed{p} \pm M)/2M$ on both sides of the bilinear. 
The different location of the $\gamma^0$ and the relativistic normalization means that the definition of positive energy states differs\footnote{ One has $\phi_{\alpha,+} - \frac{\slashed{p} + M}{2M} \tilde{\phi}_{\alpha} \simeq \frac{-\vb{p}^2}{2M^2}g + \frac{\lvert\vb{p}\rvert}{2M}f$ and $f \sim g\frac{\lvert \vb{p} \rvert}{2M}$, where $g$ and $f$ and the upper and lower components of the Dirac wavefunction cf. \cref{eq:upper-lower}.} at order $\vb{p}^2/M^2$.

A consequence of the sign changes of momenta of negative energy states is that the positive and negative energy projections are trivially orthogonal. One finds $\phi^\dagger_{\alpha,\mp}(\vb{p}) \phi_{\beta,\pm}(\vb{p}) = 0$ identically.\!\footnote{From $\tilde{\slashed{p}}^\dagger = \gamma^0\tilde{\slashed{p}}\gamma^0 = \slashed{p}$ it follows that $\phi^\dagger_{\alpha,-}(\vb{p}) \phi_{\beta,+}(\vb{p}) \propto \tilde{\phi}_{\alpha}^{\dagger} \gamma^0 \left( \slashed{p}-M \right) \left(\slashed{p}+ M \right) \gamma^0\tilde{\phi}_{\beta} = 0.$}
This means that there are no mixed contributions to the total vector density; all $\overline{\phi}_{\alpha,\mp} \gamma^0 \phi_{\alpha,\pm}$ terms disappear.
This seems intuitive, one could interpret this as $\bar{N}N$ pairs not contributing to the charge- or baryon number densities.
It is not obvious that a definition without the sign changes in momenta satisfies this property. By the same reasoning it is clear that in that case there are no mixed contributions to the scalar density instead.

\subsection{Current (non)conservation}
In the above construction we have shown how to take an operator-valued definition of the current, as in \cref{current-def}, and consistently relate free-particle matrix elements to matrix elements of particles bound in external vector or scalar fields.  Our discussion has not derived the currents given above from a microscopic model (i.e., chiral perturbation theory), and we have simply taken the simplest {\it ansatz} for their construction. As already discussed, there are other operator valued definitions of $\hat{J}_\mu^{(1)}$ that can reproduce all on-shell matrix elements. In this sense the on-shell matrix elements  are not sufficient to supply  unique operator level model-definition. As we will now see, the naive definition in \cref{current-def} leads to amplitudes which violate Ward identities (or equivalently current conservation). 

If one has an elementary current, such as $J_\mu = \bar{N}\gamma_\mu N$ (i.e.,without form factors), then the current can be easily shown to be conserved. This is most easily achieved by using the representation of the matrix elements given in \cref{elementary-current}, and using the Dirac equation \cref{Dirac-equation-phi-alpha}. 
This is also the case for the elementary current of \cref{eq:Hint_elementary}, since the anomalous magnetic moment term is constructed explicitly using derivatives of the photon field.

When considering a current defined with form factors, however, one does not obtain the simple structure of the matrix element given in \cref{elementary-current}, and the resulting current's matrix elements do not automatically satisfy the relevant Ward identities, 
\begin{equation}
    q^\mu \mel{\beta}{\hat{J}_\mu}{\alpha} \neq 0 ~,
\end{equation}
where 
\begin{equation}
    q^\mu = (\omega, \vb{q})=(E_\beta-E_\alpha, \vb{p}_\beta-\vb{p}_\alpha)~.
\end{equation}
This may be ascribed to the mismatch between the energy transfer to the nucleus $E_\beta-E_\alpha$ and the energy that would be transferred to a free-nucleon $E(\vb{p}')-E(\vb{p})$. Our discussion on elementary current in \cref{sec:elementary_my_dear} sharpens this notion however. One sees that there is no intrinsic issue with current conservation related to the on-shell treatment in that case. Only once form factors are introduced do issues arise.

It is not surprising that the nuclear model that we imposed (by fiat) does not satisfy the relevant Ward identities. The current we have lifted from free-nucleon matrix elements must satisfy $-\iu [\hat{H}_0,\hat{\rho}]= \vb*{\nabla}\cdot \vb{J}$ with $\hat{H}_0$ the free Hamiltonian. However when we add background (binding) fields $\hat{H}_0 \rightarrow \hat{H}_0 + \hat{V}$ with $\hat{V}$ a bilinear in the fields. Therefore the currents themselves must depend on the background fields (see also Ref.~\cite{Brown:1966zza} for a more general discussion) that give rise to the bound states in the mean field model. This is well known in studies of the chiral Lagrangian and electromagnetism \cite{Pastore:2008ui,Pastore:2009is,Pastore:2011ip} where the current and binding Hamiltonian must be mutually consistent in order to satisfy the continuity equation. 

At a more pedestrian level, this can also be understood diagrammatically in models where form factors arise from loops of dynamical fields. There are diagrams where the background fields attach to the external legs (outside the loop); these are resummed using mean-field wavefunctions. There are also, however, diagrams in which the background fields insert {\it inside} the loop which generates the form factor. Thus the currents will, in general, depend on the background fields and this dependence can never be inferred from free-nucleon matrix elements.
In such a microscopic model, such as chiral perturbation theory, one can compute explicitly the necessary in-medium modifications of form factors from first principles such that the Ward identities are satisfied \cite{Pastore:2008ui,Pastore:2009is,Pastore:2011ip}.

\section{Phenomenology \label{sec:phenomenology} } 
We now turn to phenomenological applications, with an emphasis on practical implications of the ``explicitly on-shell'' treatment outlined above.
We begin with elastic electron scattering, where we discuss the impact of space-like vs. time-like form factors. 
One finds that the relative size of the different contributions in \cref{main-result} follow a clearly hierarchical pattern with the non-relativistic terms dominating. 
We include time-like form factors explicitly in this case, which reduces negative energy contributions further. 

Next, we discuss pion photo-production on a nucleus. We compute the leading $++$ result, including relativistic corrections and spin-flip contributions. We show that large arbitrary uncertainties are obtained when using \cref{prescription}, and the way in which this issue is remedied in the explicitly on-shell treatment. 

\subsection{Elastic scattering \label{sec:elastic} }
Let us consider elastic scattering of electrons from a bound nuclear system, $A$, 
\begin{equation}
    e(\vb{p}) + A(\vb*{0}) \rightarrow e(\vb{p}-\vb{q}) + A(\vb{q})~. 
\end{equation}
Assuming single-photon exchange the amplitude for this process is given by the contraction of lepton and hadron vector currents $l_\mu J^\mu$.
The hadron current can be constructed from the independent four momenta $P_A^\mu, q^\mu$, where $q^\mu = (-q^2/(2M_A), \vb{q})$ is the four-momentum transfer to the nucleus.
That is $J^\mu = BP_A^\mu + B^\prime q^\mu$, where $B,B^\prime$ are functions of the invariants $Q^2\equiv -q^2, M_A$. 
These functions are not independent since current conservation requires $q\cdot J = (B/2 - B^\prime)Q^2 =0$. 
In any case the amplitude $B^\prime$ does not contribute since the lepton current is conserved $q\cdot l = 0$.
Writing $B= F_{\rm ch}(Q^2)/M_A$, the contraction of lepton and hadron currents may be readily evaluated in the lab frame
\begin{equation}
    l_\mu J^\mu = l_0J^0 = l_0 F_{\rm ch}(Q^2)~.
\end{equation}
In the following we will neglect nuclear recoil when evaluating the hadron current, i.e. treat the energy transfer $\omega \equiv q^0 = Q^2/(2M_A) \simeq 0$. The use of a central potential means that $M_A$ should be larger than any single-particle energy scale. 
Current conservation then requires that the current has no spatial components in the lab frame.\!\footnote{The spatial components are $\hat{\vb{q}}\cdot \vb{J} = \lvert \vb{q}\rvert /(2M_A) F_{\rm ch}$. The frame where the current truly has no spatial components is the Breit frame where $\vb{p}_A = -\vb{p}_A^\prime$, and $J^0 =F_{\rm ch}(Q^2) \sqrt{1 + \vb{p}_A^2/M_A^2}$, this frame is practically the lab frame frame, the velocity of the Breit frame in the lab frame is $\vb*{\beta} = \frac{\vb{q}}{2M_A + \omega}$}

We consider the one-body current contribution to the hadron current. By \cref{current-def} it may be decomposed as
\begin{equation}
\label{eq:J0_decomp}
    J^0 = J^0_{++} +J^0_{--}+J^0_{-+}+J^0_{+-},
\end{equation}
where the subscripts denote positive and negative energy contributions.
These may be written in terms of bilinears of positive/negative energy wavefunctions defined in \cref{sec:Currents_w_FF}.
Using the notation of \cref{main-result} we have
\begin{align}
    &\Gamma_\mu^{(a)}(\vb{p}^\prime,\vb{p}) = \Gamma_{\mu}^{(b)}(-\vb{p}^\prime,-\vb{p}) = F_1( P^2)\gamma_\mu + \iu\frac{F_2(P^2)}{2M_N} \sigma_{\mu\nu} q^\nu, \label{eq:gammaa_elastic} \\
    &\Gamma_\mu^{(c)}(-\vb{p}^\prime,\vb{p}) = \Gamma_{\mu}^{(d)}(\vb{p}^\prime,-\vb{p}) = F_1( \tilde{P}^2)\gamma_\mu + \iu\frac{F_2(\tilde{P}^2)}{2M_N} \sigma_{\mu\nu} q^\nu, \label{eq:gammac_elastic}
\end{align}
where $\vb{p}^\prime = \vb{p} + \vb{q}$.
The sole difference is the point at which the Dirac and Pauli form factors, $F_1$ and $F_2$ respectively, are evaluated:
\begin{align}
    P^2 &\equiv -(p - p^\prime)^2 
    =  \vb{q}^2 - (E(\vb{p}) - E(\vb{p}^\prime) )^2, \\
    \tilde{P}^2 &\equiv -(p + \tilde{p}^\prime)^2 
    =  \vb{q}^2 - (E(\vb{p}) + E(\vb{p}^\prime))^2,
\end{align}
with $E(\vb{p}) =\sqrt{ \vb{p}^2 + M_N^2}$ and $\vb{p}'=\vb{p}+\vb{q}$.
We have identified $q^\mu$ with the four-momentum of the virtual photon for both cases in analogy with \cref{eq:Hint_elementary}. This does not affect the time-like component of the current since $\sigma^{0\nu}q_\nu \propto \lvert \vb{q} \rvert$ in any case.
In \cref{app:useful-formulae}, we give convenient expressions for the positive/negative energy projections of the wavefunctions that enter in \cref{main-result}, and derive results for coherent interactions with closed shell nuclei for which the sum over angular momenta is performed analytically.  
Using these results, explicit expressions for the 4 contributions to the current of \cref{eq:J0_decomp} are given in \cref{eq:J0ssprime_result,eq:Jdens_result}.
Note that one finds $J^0_{+-} = J^0_{-+}$.

\begin{figure}
    \centering  \includegraphics[width=\linewidth]{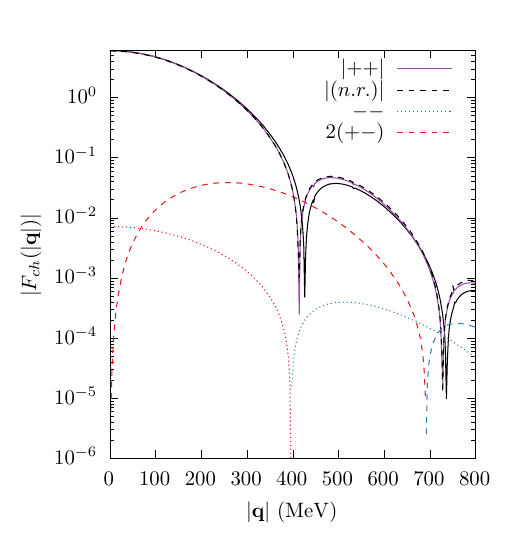}
    \caption{Contributions to the total charge form factor of ${}^{12}$C (black lines) where form factors from~\cite{Kelly04} are included in the space like region at fixed values of $Q^2=\vb{q}^2$. For the $--$ and $+-$ contributions, the sign is indicated by the color of the curves with red(blue) corresponding to positive (negative). Also shown is the non relativistic approximation.}
    \label{fig:F_ch_log}
\end{figure}

To make a connection to the existing literature, let us consider the same bilinear for each contribution $(a,b,c,d)$: 
\begin{equation}
\label{eq:Gamma_gi}
    \Gamma^\mu = g_1\gamma^\mu + \iu \frac{g_2}{2M} \sigma^{\mu\nu}Q_\nu,
\end{equation}
where $g_1, g_2$ are constant, e.g. $g_i = F_i(\vb{q}^2$), the form factors at fixed momentum transfer. 
We will discuss corrections to this result from the inclusion of the momentum dependence of the form factors. 
In this case, since there is no explicit momentum dependence, \cref{elementary-current} holds. 
Inserting the Fourier transforms of wavefunctions, the current (again for closed shell nuclei) is readily obtained from the coordinate space wavefunctions 
\begin{align}
    J^0 &= \sum_{\kappa,m} \int \mathrm{d}\vb{r} \e^{-\iu\vb{q}\cdot\vb{r}} \overline{\Psi}_\kappa^{m}(\vb{r}) \Gamma^0 \Psi_\kappa^m(\vb{r}) \\ &= \int \mathrm{d}\vb{r} \e^{-\iu\vb{q}\cdot\vb{r}} \mathrm{Tr}\left\{ \Gamma^0 \left[ \sum_{\kappa,m} \Psi_\kappa^m(\vb{r}) \overline{\Psi}_\kappa^{m}(\vb{r}) \right] \right\} \label{eq:spinsum_explicit} \\
    &= \int \mathrm{d}\vb{r} \e^{-\iu\vb{q}\cdot\vb{r}} \frac{1}{8\pi} \mathrm{Tr}\left\{ \Gamma^0 \left[ \gamma^0 \rho^V + \rho^S + \iu \gamma^0\hat{\vb{r}}_i\gamma^i \rho^T \right] \right\} \label{eq:spinsum_perfomed} \\
    &= \int r^2\mathrm{d}r \left[ j_0(qr) g_1 \rho^V(r) + j_1(qr)  \frac{ g_2 \lvert\vb{q} \rvert}{2M_N}\rho^T(r) \right].
\end{align}
In going from \cref{eq:spinsum_explicit} to \cref{eq:spinsum_perfomed} the sum over angular momenta $m$ is performed in the same way as in \cref{app:useful-formulae}, and written in a convenient way in terms of scalar, vector and tensor densities~\cite{Gardner:1994bh}.
The current is given by Hankel transforms of the vector and tensor densities~\cite{PhysRevC.86.045503}
\begin{equation}
    \frac{\rho^V(r)}{2j+1} = \sum_\kappa f^2_\kappa(r) + g_\kappa^2(r),\quad \frac{\rho^T(r)}{2j+1}=\sum_\kappa 2f_\kappa(r)g_\kappa(r).
\end{equation}
The current may again be decomposed into the four contributions of \cref{eq:J0_decomp}, given in \cref{app:elastic_scattering}.
These are shown for ${}^{12}$C in \cref{fig:F_ch_log}. Calculations are performed with the model of Ref.~\cite{Sharma93}, we only include protons.

Following the discussion in \cref{sec:Currents_w_FF} we may isolate the leading term $J_{++}^0$ and relate it to a non-relativistic treatment.
The single nucleon amplitude may be written in terms of the bilinear operator $\mathcal{F}^0_{++}$ defined\footnote{In short: $\chi^\prime \mathcal{F}^\mu_{++}(\vb{p}^\prime,\vb{p}) \chi \equiv \overline{u}(\vb{p}^\prime) \Gamma^\mu(\vb{p},\vb{p}^\prime) u(\vb{p})$ where $\chi^{(\prime)}$ are two-component Pauli spinors.} in \cref{eq:defmathFPauli}.
For \cref{eq:Gamma_gi} one finds
\begin{align}
    \mathcal{F}^0_{++}(\vb{p}^\prime,\vb{p}) &=  g_1\left[1+\frac{(\vb*{\sigma}\cdot\vb{p}^\prime)(\vb*{\sigma}\cdot\vb{p})}{(E^\prime+M)(E+M)}\right] \\ &+ g_2\left[ \frac{(\vb*{\sigma}\cdot \vb{q})(\vb*{\sigma}\cdot \vb{p})}{2M(E+M)} - \frac{(\vb*{\sigma}\cdot \vb{p}^\prime) (\vb*{\sigma}\cdot \vb{q})}{2M(E^\prime+M)}\right].
\end{align}
The non-relativistic reduction is
\begin{align}
\mathcal{F}^0_{++}(\vb{p}^\prime,\vb{p}) & \simeq g_1\left[ 1 + \frac{ \vb{p}\cdot\vb{p}^\prime}{4M^2} +\iu\frac{\vb*{\sigma}\cdot(\vb{p}^\prime \cross \vb{p})}{4M^2} \right] \nonumber \\
&+ \frac{g_2}{4M^2} \left[ -\vb{q}^2 + 2i\vb*{\sigma}\cdot(\vb{q}\cross \vb{p}) \right] + O\qty(\frac{\vb{p}^4}{M^4}) ,
\label{eq:Felastic_nonrel}
\end{align}
Note that to obtain the Darwin  term  in its usual form ($\vb{p}\cdot\vb{p}^\prime/4M^2 \rightarrow -\vb{q}^2/8M^2$)  one should include the normalization factor of the Dirac spinors~\cite{PhysRevA.56.4579,PhysRev.78.29}, which we have absorbed in the wave functions.\!\footnote{The scalar(spin-dependent) terms then yield the $\mathcal{F}^M (\mathcal{F}^{\phi^{\prime\prime}})$ responses defined in~\cite{Hoferichter:2020osn}. Using the notation of \cref{app:elastic_scattering}, for a closed shell with angular momenta $\kappa = \{J,l\}$ these are 
\begin{equation}
\mathcal{F}^{M} \propto (2J+1)\int \mathrm{d}^3\vb{p}~\tilde{g}_{\kappa,+}(\lvert \vb{p}^\prime\rvert) \tilde{g}_{\kappa,+}(\lvert \vb{p}\rvert) P_l(\hat{\vb{p}}^\prime\cdot\hat{\vb{p}}), \nonumber
\end{equation}
\begin{equation}
    \mathcal{F}^{\Phi^{\prime\prime}} \propto \frac{2}{\vb{q}^2}\int \mathrm{d}^3\vb{p} \tilde{g}_{\kappa,+}(\lvert \vb{p}^\prime\rvert)\tilde{g}_{\kappa,+}(\lvert \vb{p}\rvert) \frac{\lvert \vb{p}^\prime \cross \vb{p} \rvert^2 }{\lvert \vb{p}^\prime \rvert \lvert\vb{p}\rvert} P^\prime_l(\hat{\vb{p}}^\prime\cdot\hat{\vb{p}}). \nonumber
\end{equation}
}
In any case this doesn't affect our main conclusions. The result $J_{(n.r)}^0$ obtained with this approximation is included in \cref{fig:F_ch_log}.
Clearly, it reproduces the $J_{++}^0$ result well. This is expected for coherent interactions since $p/M$ is a suitable small parameter.\!\footnote{Note that we did not renormalize the density obtained from the upper components of the wavefunctions to one, which would be the case in a non-relativistic model}
The complete result for $\lvert J_0 \rvert$ computed in the RMF with form factors fixed to spacelike values does differ markedly from $J^0_{++}$. 
The position of the zero is shifted to larger $\lvert \vb{q} \rvert$.
This is almost completely due to the $g_2$ contribution to $J_{+-}^0 = J^0_{-+}$.
This is a relativistic effect: the $g_2$ contribution mixes upper and lower components of the spinors, and the latter are enhanced due to the potential.
However, as we will show, when the form factors that enter in $J^0_{\pm,\mp}$ are evaluated at timelike momenta, the contribution of this term becomes an order of magnitude smaller and one finds results consistent with only retaining $J_{++}^0$.

We now consider the momentum dependence of the form factors.
For $J_{++}^0$ and $J_{--}^0$ the form factors enter in the timelike region $P^2 = \vb{q}^2 -(E^\prime-E)^2 \simeq \vb{q}^2$, in \cref{eq:gammaa_elastic}.
Using the momentum-space formalism, the full dependence can be readily included.
Such corrections are small, and may be included by Taylor expansion of the form factors around the fixed value $\vb{q}^2$.
If we assume per illustration a dipole form $F_i(x) \propto (1+x/M_D^2)^{-2}$, the Taylor expansion is 
\begin{equation}
    F_i(P^2) = F_i(\vb{q}^2)\left[ 1 + \frac{2}{(1+\vb{q}^2/M_D^2)} \frac{(E^\prime - E)^2}{M_D^2} \right],
\end{equation}
and since $(E^\prime - E)^2 \simeq (2\vb{p}\cdot\vb{q} + \vb{q}^2)^2/(4M^2)$ and $M_D \sim M$, these corrections are $p^4/M^4$ for the leading $F_1$ contribution in $J^0_{++}$. 
Hence, inclusion of the momentum dependence of the form factor is essentially irrelevant for the $J^0_{\pm\pm}$ contributions.

\begin{figure}
    \centering
    \includegraphics[width=\linewidth]{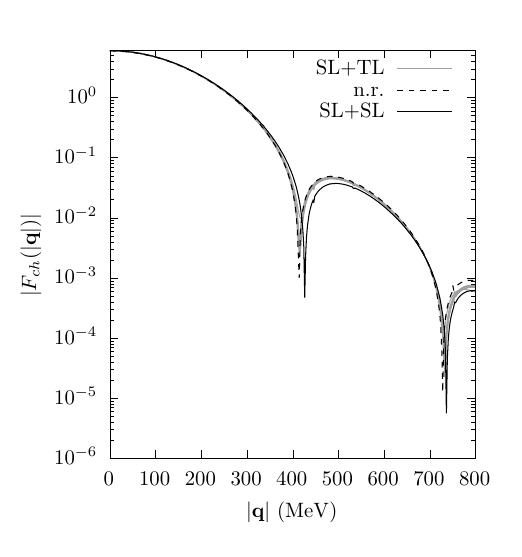}
    \caption{Charge form factor, the full results (SL+TL) where nucleon form factors in the timelike region are included in the $\pm,\mp$ terms, obtained using 7 different parametrizations, are shown by grey lines. The non-relativistic approximation $(n.r.)$, and the result where all form factors are evaluated in the spacelike region (SL+SL) are shown by dashed and solid black lines respectively.}
    \label{fig:F_ch_SLTL}
\end{figure}
The interesting case are the mixed contributions $J^0_{\pm\mp}$ where the form factors enter at time-like four-momenta 
\begin{equation}
    \tilde{P}^2 = \vb{q}^2 - (E^\prime + E)^2 \simeq -4M_N^2 - (2\vb{p} + \vb{q})^2 + O\qty(p^2/M^2)
\end{equation}
In this case, the momentum dependence around the central value is larger, but we will neglect the $\vb{p}$ dependence, assuming that the slope of the form factor in the time like region is again $1/M^2$, and that such dependence is smaller than uncertainties in the form factors. We focus on the more significant result that the form factors are evaluated in the time-like region.

Nucleon form factors in the timelike region are far less constrained than those in the spacelike regime~\cite{Denig:2012by}.
Our goal is not to perform a full analysis of the sensitivity to the time like form factors.
Instead we consider some readily available parametrizations for illustration; all are based on vector meson dominance.
These are the model of Refs.~\cite{Iachello:2004aq, IACHELLO1973191, Tomasi-Gustafsson:2005svz}, using the parametrization given in~\cite{Tomasi-Gustafsson:2005svz} with the two parameter sets given in Table V of that reference.
The model of~\cite{Lomon:2012pn}, using the four reported parameter sets.\!\footnote{In implementation of this model we modified the reported value of the pole position in the $\omega^\prime$ contribution to the correct value, $s_s \approx 0.146~\mathrm{GeV}$, and changed the sign of $\kappa_{\rho^\prime}$ for 'case 1 with Babar'.}
And the model of Ref.~\cite{Yan:2023nlb}, where we use the reported central values of the parameters.

 In \cref{fig:F_ch_SLTL} we show the result where now form factors in the $\pm,\mp$ contributions are included at a fixed value of $Q^2 = -\vb{q}^2 - (E(\vb{q}) + M )^2$ in the time-like region.  
One sees that the result is closer to the non-relativistic approximation in this case.
The reason is that the value of $F_2$ tends to be about an order of magnitude smaller in the time-like region than in the spacelike region, so that the $J^0_{+-}$ contribution is reduced.
The enhancement of $F_2$ in the $+-$ current from the large lower components in the RMF, is essentially nullified by the fully relativistic treatment where form factors are evaluated in the time like region.

Lastly, we examine current conservation.
When one uses the bilinear of \cref{eq:Gamma_gi} with form factors constant, the current is conserved when recoil is neglected, in this case: $\vb{q}\cdot \vb{J}=0$. 
For the $g_2$ term we used a fixed $Q=(0,\vb{q}$), such that this contribution to the current is trivially conserved.
The $\gamma^\mu$ contribution is conserved when the wave functions are solutions of the Dirac equation in the same potential, which may be derived from \cref{elementary-current}.
When different couplings enter in the $\pm\pm$ and $\pm\mp$ contributions, the cancellation between terms used to derive \cref{elementary-current} from \cref{main-result} does not occur, and it is not obvious that the current is conserved.
However, as shown in \cref{app:elastic_scattering}, for elastic scattering a cancellation occurs between the mixed contributions, $\vb{q}\cdot \vb{J}_{+-}=-\vb{q}\cdot\vb{J}_{-+}$. Since the full current is conserved the mixed and diagonal contributions are thus separately conserved
\begin{equation} 
\label{eq:longitudinal_cancellation_elastic}
\vb{q}\cdot(\vb{J}_{+-}+\vb{J}_{-+})=\vb{q}\cdot(\vb{J}_{++} + \vb{J}_{--})=0.
\end{equation}
This means one can use different couplings for these diagonal and mixed contributions without spoiling current conservation. Moreover, since as discussed in \cref{sec:Currents_w_FF} $J_{\pm,\mp}^0(\vb{q} = \vb{0}) = 0$, these different couplings do not affect the normalization.
When the full momentum dependence of the form factors is included, this is no longer the case, and one will violate current conservation at the order $\sim p^4/M^4$.
These violations of current conservation due to the momentum dependence of the form factors
point to the true deeper lying issue for a consistent nuclear amplitudes: the  lack of a fundamental  description of form factors, which must depend on the mean-field potentials for a conserved vector current.

\begin{figure*}
    \centering
    \includegraphics{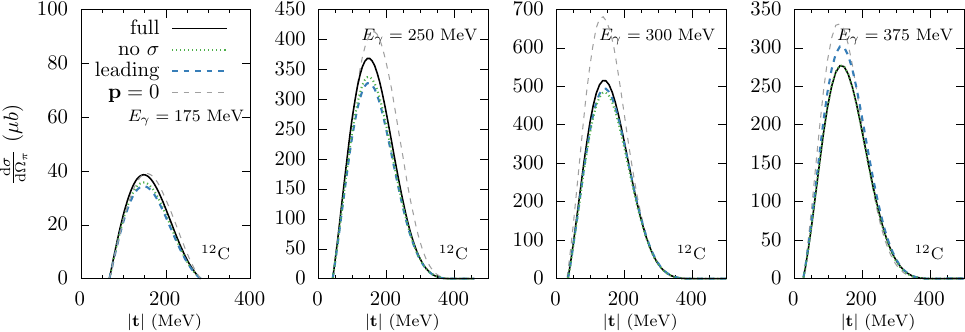} \\
    \includegraphics{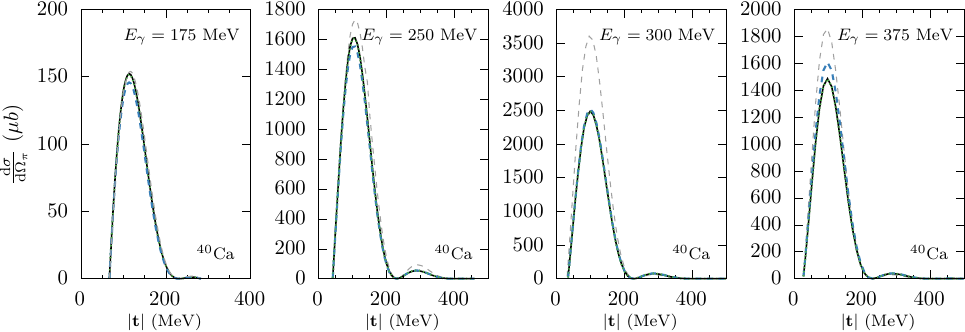}
    \caption{PWIA results for coherent pion photoproduction off ${}^{12}$C  (top) and ${}^{40}$Ca at different photon energies as function of the momentum transfer.
    The result labeled no $\sigma$ neglects the spin-flip contributions. The 'leading' result is defined in \cref{eq:F0_leading}. The result labeled $\vb{p} = 0$ is the 'leading' contribution but where the $A_1$ invariant amplitude is evaluated at a fixed value of $\vb{p}=0$.}
    \label{fig:CS_coherent_diff}
\end{figure*}

\subsection{Coherent pion photo-production \label{sec:pion-photo-prod}} 
Pion production is an important topic in hadron physics. It provides us access to the axial current, and is an important detection channel in neutrino physics~\cite{NUSTECWP}. Pion photoproduction is remarkably well characterized on free nucleons~\cite{PhysRevC.108.065205, Anisovich:2010wx}, and is an instructive example to study to better understand the limitations and merits of nuclear mean field models. The wealth of experimental data means that many modeling assumptions can be compared to data. 

Relativistic mean field treatments have found a disturbing level of sensitivity to the arbitrary choice of parametrization of the hadronic one-body currents. 
In particular the authors of Ref.~\cite{Abu-Raddad:1998ams} conclude that different choices of the basis for invariant amplitudes that parametrize the free-nucleon process can change the resulting nuclear-level amplitude for pseudoscalar meson production by factors as much as $500\%$. 
With the results derived in this work, that is Eq.~(\ref{main-result}), these arbitrary choices give unique results. 
In particular, it allows to isolate the $++$ contribution which should provide a reliable and systematic leading order approximation of the full mean-field result. 
The main uncertainty comes from the $\pm,\mp$ contributions, for which one now obtains unique results. Our treatment allows to estimate and parametrize their contribution in a systematic way.

The process under consideration is
\begin{equation}
    \gamma(\vb{q}) + A(\vb*{0}) \rightarrow A(\vb{t} = \vb{k}_\pi - \vb{q}) + \pi(\vb{k}_\pi).
\end{equation}
The cross section in the target rest frame may be written as
\begin{equation}
    \frac{\mathrm{d}\sigma}{\mathrm{d}\Omega_\pi} = \frac{1}{16\pi^2} \frac{M_A}{E_A^\prime f_{rec}} \frac{\lvert \vb{k}_\pi\rvert}{\lvert \vb{q} \rvert} \frac{1}{2}\sum_{\lambda} \lvert \epsilon_\lambda \cdot J \rvert^2,
\end{equation}
where $f_{rec} = \lvert 1 + E_\pi/E_{A}^\prime \left( 1 - \vb{q}\cdot\vb{k}_\pi/\vb{k}_\pi^2 \right) \rvert$.
Since the nuclear current is a pseudovector built from the relevant four-momenta, one only has one option:
\begin{equation}
    J^\mu = \frac{A(s,t)}{M_A}~\epsilon^{\mu\alpha\beta\gamma}P_{A,\alpha} q_\beta k_{\pi,\gamma},
\end{equation}
where $A$ is a scalar function of the invariants $s = (P_A + q)^2$ and $t= (q - k_\pi)^2$. In the target rest frame one has
\begin{equation}
\label{eq:generalJmu}
    \epsilon \cdot J = A(s,t) \vb*{\epsilon}\cdot (\vb{q}\cross \vb{k}_\pi),
\end{equation}
and only components of the polarization vector orthogonal to $\vb{k}_\pi$ contribute. Hence the sum over polarizations may be dropped if one takes $\vb*{\epsilon} = \widehat{\vb{q}\cross\vb{k}_\pi}$. All current (operators) in the following without an explicit Lorentz index are implicitly contracted with this vector.

The one body current contribution to $\epsilon\cdot J$ from a single-particle state labeled with quantum numbers $\alpha$ may be written as in Eq.~(\ref{main-result})
\begin{equation}
    \label{++-def}
    \begin{split}
    \langle \alpha \pi \lvert \hat{J}^{(1)} \lvert \alpha \rangle \simeq
    &\int_{\vb{p}}
    \overline{\psi}_{\alpha,+}(\vb{p}^\prime) \Gamma^{(a)}(\vb{p}^\prime, \vb{p}, \vb{k}_\pi^\prime) \psi_{\alpha,+}(\vb{p}) \\
    +&\int_{\vb{p}} 
    \overline{\psi}_{\alpha,-}(\vb{p}^\prime) \Gamma^{(b)}(-\vb{p}^\prime, -\vb{p}, \vb{k}_\pi^\prime) \psi_{\alpha,-}(\vb{p}) \\
    + &\ldots
    \end{split}
\end{equation}
where $\int_{\vb{p}} = \int \dd^3p/(2\pi)^3$ and the pion momentum is fixed because we use the plane-wave impulse approximation for the pions. In a distorted-wave approach these are the matrix elements which should be folded with the pion wavefunction. 
The states $\psi_{\alpha,\pm}$ are the appropriate positive/negative energy projections given explicitly in \cref{app:useful-formulae}, and $\vb{p}^\prime = \vb{q} -\vb{k}_\pi + \vb{p}$.

The bilinear operators $\Gamma^{(a,b,c,d)}$, are defined again by \cref{eq:def_bilinears}, but where now a pion is added in the final-state. 
Matrix elements for single pion production can be parametrized, in the same way as the single nucleon current, in terms of a set of invariant amplitudes as (see \cref{app:cohpi} for more details)
\begin{equation}
    \Gamma^\mu = \sum_{i=1}^4 A_i(s,t,u) \mathcal{M}_i^\mu,
\end{equation}
Where the amplitudes are defined by the choice $\mathcal{M}_i^\mu$, and are scalar functions of the invariants.
A set of bilinears considered in Refs.~\cite{Abu-Raddad:1998ams, BENNHOLD1991625} is
\begin{align}
\label{eq:M1_def}
    &\mathcal{M}_1^\mu = -\gamma^5 \gamma^\mu \slashed{q} ~,\\
    &\mathcal{M}_2^\mu = 2\gamma^5 \left[ p^\mu ( q\cdot p^\prime) - p^{\prime,\mu}(q\cdot p )\right]~, \\
    &\mathcal{M}_3^\mu = \gamma^5 \left[\gamma^\mu (q\cdot p) - \slashed{q}p^\mu \right]~, \\
    &\mathcal{M}_4^\mu = \gamma^5 \left[\gamma^\mu (q\cdot p^\prime) - \slashed{q}p^{\prime,\mu} \right].
    \label{eq:M4_def}
\end{align}
This parametrization gives the $\Gamma^{(a)}$ term.
The Dirac structure of $\Gamma^{(b,c,d)}$ may be taken to be the same (with appropriate sign changes in the four momenta) since these processes are related by crossing~\cite{BERENDS19671, Chew:1957tf}.
We will not attempt to construct them here as they would require time-like data from e.g., $e^+e^- \rightarrow N\bar{N} \pi$. 

The bilinears depend on the momenta $\vb{p},\vb{p}^\prime$, and the condition of~\cref{eq:condition} will typically not be met. 
If one uses this (or any equivalent) form of $\Gamma^{(a)}$ in~\cref{elementary-current}, spurious contributions will be present in the $(b,c,d)$ contributions, completely analogous to the discussion in~\cref{sec:elementary_my_dear} which we will illustrate below.

\begin{figure}
    \centering
    \includegraphics[width=0.48\textwidth]{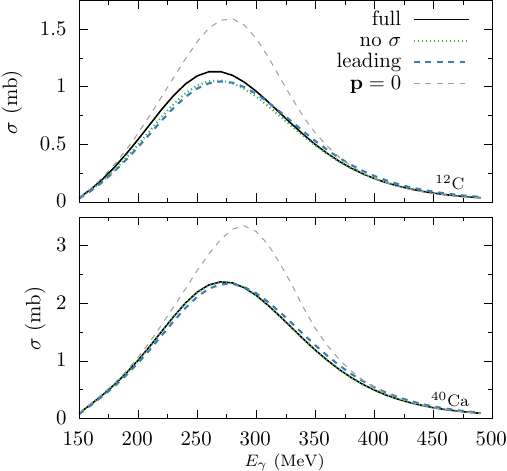}
    \caption{Total PWIA cross sections for ${}^{12}$C (top) and ${}^{40}$Ca (bottom). The labels are the same as in \cref{fig:CS_coherent_diff}.}
    \label{fig:CS_tot_SPP_pp}
\end{figure}

In the on-shell treatment any parametrization of $\Gamma^{(a)}$ that gives the same free-nucleon amplitudes for pion electroproduction\footnote{Any two sets of amplitudes that include
longitudinal contributions and are related by using energy momentum conservation and the free Dirac equation but \emph{not} $q^2=0$.} yields the same nuclear $++$ amplitudes.
As explained in \cref{sec:Currents_w_FF}, this should be the leading contribution, which may be matched to ``traditional'' non-relativistic treatments, we discuss it in some detail first.

\Cref{fig:CS_coherent_diff} shows the differential section, while \cref{fig:CS_tot_SPP_pp} shows the total cross section where only positive energy states are retained.
In these computations the $A_i$ are obtained from the ANL-Osaka model~\cite{Kamano2013,DCC:electron}, by relating these to multipole amplitudes, obtained from~\cite{Kamano:2019gtm}, as described in \cref{app:CGLN}. The nuclear wavefunctions are computed with the model of Ref.~\cite{Sharma93}.
We always use the on-shell treatment $q^0=E(\vb{p}^\prime ) +E( k_\pi ) - E(p)$. For simplicity, we did not include the explicit $q^2$ dependence of the amplitudes however. 
We treat the proton/neutron contributions explicitly by using the appropriate amplitudes for $\pi^0$ production in either case. Of course for symmetric nuclei (neglecting Coulomb corrections) one can use the isovector amplitudes instead.

One may write the single nucleon amplitude in two-component form in terms of the operator
\begin{equation}
\label{eq:F_paulivecF0}
   \vb*{\epsilon}\cdot\vb*{\mathcal{F}}_{++} = F_0(\vb{p}, \vb{q},\vb{t}) + \iu \vb*{\sigma}\cdot\vb{K}(\vb{p}, \vb{q},\vb{t}),
\end{equation}
where we omitted Pauli-spinors.
The explicit expression for the functions $F_0$ and $\vb{K}$ are given in~\cref{eq:edotJ_Aamps}.
The $F_0$ term leads to the coherently enhanced part of the nuclear amplitude.
The spin-dependent $\vb*{\sigma}\cdot\vb{K}$ is suppressed, it contributes due to the orbital angular momentum of the bound nucleons.
As expected, the spin-flip contribution is more relevant for smaller nuclei, and becomes irrelevant for calcium. Note that in calcium there is of course a cancellation between the filled $p_{1/2}, p_{3/2}$ and $d_{3/2}, d_{5/2}$ states. These pairs contribute with opposite sign but similar radial wave functions.

A naive\footnote{Assuming that $A_1 \sim A_3/M_N \sim A_4/M_N$ based on the mass dimensions of the associated $\mathcal{M}_i$ and $\lvert \vb{p} \rvert \sim \lvert \vb{t} \rvert \sim \lvert \vb{q} \rvert$} power counting of \cref{eq:edotJ_Aamps} allows to identify the leading contribution as
\begin{equation}
\label{eq:F0_leading}
    F_0 \simeq \frac{A_1}{(E^\prime+M_N)} \vb*{\epsilon}\cdot ( \vb{q} \cross \vb{t}).
\end{equation}
Other terms are suppressed by powers of $|\vb{t}|/M_N,~ |\vb{q}|/M_N,~|\vb{p}|/M_N$ in comparison.
The cross sections obtained by only including this leading term are also shown explicitly in \cref{fig:CS_coherent_diff,fig:CS_tot_SPP_pp}.
This is clearly a good approximation in the kinematic region under consideration. For ${}^{12}$C, neglecting the spin-dependent terms is seen to have larger effects.

This identification of a 'leading term' of course depends on the choice of basis.
For example, as sketched in  \cref{app:CGLN}, the single-nucleon amplitude may be expressed directly in terms of the CGLN amplitudes $F_i$~\cite{Chew:1957tf}.
In this case the leading term (obtained by neglecting corrections from boosts to the CMS) is given entirely by the $F_2$ CGLN amplitude.
From \cref{eq:A1_cgln} one sees that $A_1$ is indeed dominated by the $F_2$ amplitude, with a highly suppressed contribution from $F_1$.
This is understood as the result in the static nucleon limit (where $\vb{p}= \vb{0}$).
The suppressed $F_1$ contribution then stems from the Wigner rotation of the nucleon spin as described in \cref{app:CGLN}.

The leading contribution of \cref{eq:F0_leading}, corresponds to taking the $\vb{p}=\vb{0}$ limit of \cref{eq:edotJ_Aamps}.
The amplitude $A_1$ does however depend on $\vb{p}$, which is included in the integrals to compute the 'leading' result shown in the figures.
If this dependence can be omitted, e.g. by evaluating the amplitude at some fixed momentum (i.e. a 'local approximation'), the computation simplifies to the multiplication of the single-nucleon amplitude with a nuclear form factor. We show the result where the amplitude is evaluated at $\vb{p} = \vb{0}$ in \cref{fig:CS_coherent_diff} and \cref{fig:CS_tot_SPP_pp}. This approximation leads to a significant overestimation of the cross section near the delta resonance peak.
Note that when such approximations have been used in other works, the single-nucleon amplitude is not evaluated at $\vb{p}=\vb{0}$ but at kinematics shown to yield a more suitable approximation~\cite{PhysRevC.30.989, TIATOR1980343, Tsaran:2024sue, Zhang:2012xi, PhysRevC.19.142, CARRASCO1993797}.

Let us now illustrate what happens when one does not treat the negative energy contributions consistently using \cref{main-result}. 
That is when \cref{elementary-current} is used erroneously, or put equivalently: when $\Gamma^{(a)}(\vb{p},\vb{p}^\prime, \vb{k}_\pi)$ is inserted in each of the terms of \cref{main-result} instead of the $\Gamma^{(b,c,d)}$ with appropriate sign changes.  
One then obtains different results when equivalent parametrizations of $\Gamma^{(a)}$ are used.
We illustrate this here.

The nuclear target amplitude may be split up as
\begin{equation}
    \epsilon\cdot J = \epsilon\cdot\left[ J_{++} + J_{--} + J_{-+} + J_{+-} \right],
\end{equation}
with subscripts again denoting the positive/negative energy components of the wavefunctions.
The 'operator ambiguity' follows from the negative energy contributions.
This is clear from a simple consideration: one can multiply all (or any single one) of the bilinears $\mathcal{M}_i^\mu$ to the right by $\slashed{p}/M_N$ or to the left by $\slashed{p}^\prime/M_N$ without affecting $J_{++}$ or the definition of the invariant amplitudes.
Of course, this means the signs of the $J_{\pm,\mp}$ contributions change.
These signs are thus completely arbitrary.

\begin{figure}
    \centering
    \includegraphics[width=0.48\textwidth]{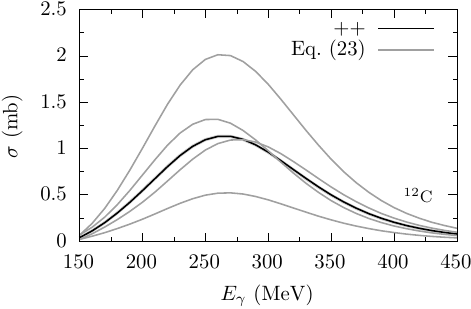}
    \caption{Total PWIA cross sections for ${}^{12}$C illustrating the 'ambiguities' due to negative energy components. The black line retains only the $++$ contribution, while the gray lines include the negative energy states. These are obtained by using different forms of $\Gamma^{(a)}$ that give the same free nucleon results in  Eq.~(\ref{elementary-current}).
    }
    \label{fig:CS_tot_ambiguities}
\end{figure}

In \cref{fig:CS_tot_ambiguities} we show results for the total cross section that include arbitrary signs in $J_{\pm,\mp}$ (there are $2^2=4$ curves corresponding to all possible binary choices). For reference, the cross section that includes only $J_{++}$ is also shown.
Results deviate by up to a factor 2. 
This is completely due to the $J_{\pm,\mp}$, the $J_{-,-}$ contribution is suppressed by an additional factor with magnitude $\sim \lvert \vb{t} \rvert/(2M)$ as expected.
We also computed the result using the '$B_i$' amplitudes\footnote{These are defined by replacing $\mathcal{M}_1^\mu \rightarrow \mathcal{M}_1^{\mu \prime} = i/(2M_N) \epsilon^{\mu\alpha\beta\gamma}q_{\alpha} (p^\prime - p)_{\beta} \gamma_\gamma$ which implies $A_3\rightarrow A_3^\prime = A_3 - A_1/(2M_N)$, $A_4\rightarrow A_4^\prime = A_4 - A_1/(2M_N)$, while $A_1^\prime = A_1$, $A_2^\prime = A_2$ remain unchanged.} of Ref.~\cite{Abu-Raddad:1998ams}. In this case the $J_{\pm,\mp}$ are strongly suppressed. The 4 different results differ at the permille level from the $++$ contributions and the curves are indistinguishable.
This explains, and supports the results in~\cite{Abu-Raddad:1998ams}, why using \cref{prescription} leads to large modeling uncertainties. In particular, the smallness of the $\pm,\mp$ terms obtained with the $B_i$ amplitudes implies that this set yields results that are essentially equivalent to retaining only positive energy states.
Of course, none of these equivalent parametrizations should be considered 'better' or 'worse', the choice is arbitrary, this issue is avoided in the explicitly on-shell treatment presented herein.

\begin{figure}
    \centering
    \includegraphics[width=0.48\textwidth]{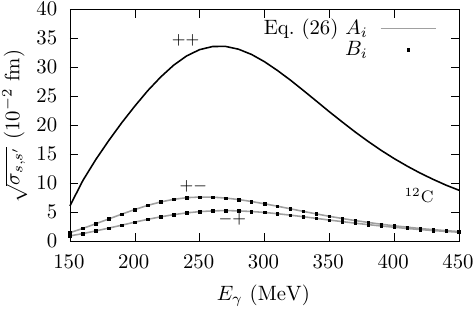}
    \caption{Square roots of the PWIA cross sections $\sigma_{s,s^\prime}$ for ${}^{12}$C. The subscripts indicate that only $++$, $-+$ and $+-$ contributions are included in the computation of the cross section. These are a measure of the magnitudes of these different contributions to the \emph{amplitude}. Results are obtained with Eq.~(\ref{main-result}) where the bilinears $\Gamma^{(c,d)}$ that enter in the $\pm,\mp$ results are defined using Eq.~(\ref{eq:signchange}). In this case one obtains unique results for different parametrizations.}
    \label{fig:CS_tot_NOambiguities}
\end{figure}

Indeed, these ambiguities do not arise when \cref{main-result} is used with well-defined $\Gamma^{(c,b,d)}$.
$J_{--}$ may be computed quite straightforwardly, since $C$-invariance relates the amplitudes for anti-nucleons to the one for nucleons~\cite{BERENDS19671, Chew:1957tf}. 
The $J_{\pm,\mp}$ currents on the other hand require matrix elements for $\gamma \rightarrow \bar{N} N \pi$ and $\bar{N}N \rightarrow \pi \gamma$.
These may be described by a bilinear constructed from the same $\mathcal{M}_i^\mu$ but with appropriate sign changes of the four-momenta~\cite{BERENDS19671}. For example, for the $-+$ contribution 
\begin{equation}
    \begin{split}
    \overline{v}(p^\prime) \Gamma^{\mu}&(\vb{p}^\prime, \vb{p},\vb{k}_\pi) u(p) \\
    =&~~\overline{v}(p^\prime) \left[\sum_i \tilde{A}_i\mathcal{M}_i^\mu(p,-p^\prime, k_\pi)\right]u(p),
    \label{eq:signchange}
    \end{split}
\end{equation}
and equivalently with $p\rightarrow-p$ for $+-$.
The set of amplitudes $\tilde{A}_i$ should now be evaluated at timelike momentum transfers.
We won't attempt to estimate the value of the amplitudes.
To illustrate the ``natural'' magnitude of $J_{\pm,\mp}$ we simply fix the values of the amplitudes to those of the direct pion production reaction.
With the sign changes of Eq.~(\ref{eq:signchange}) and those in Eq.~(\ref{main-result}) one then obtains results that are unique for different equivalent on-shell parametrizations. e.g. the $A_i$ and $B_i$.

An estimate of the relative size of these contributions is shown in the bottom panel of Fig~\ref{fig:CS_tot_NOambiguities}.
We show the square-root of the total cross section which provides an estimate of the phase-space averaged magnitude of the amplitude $\sqrt{\sigma_{s,s^\prime}} \propto \langle \lvert J_{s,s^\prime} \rvert\rangle$. The subscript indicate that only the specific ${s,s^\prime}$ contribution is included.
Note that these are large corrections at the level of the amplitude, but of course this is an upper bound since cancellations can occur.

When properly evaluated at time-like four-momenta, one expects these to be further suppressed in the same way as for elastic scattering.
 Indeed, $t \simeq 4M_N^2$ lies far above the pion threshold and beyond the mass region of the lightest vector mesons, and similarly $q^2 \simeq 4M_N^2$ should strongly reduce coupling to resonances. There is a clear delta contribution in the $\pm,\mp$ results in Fig.~\ref{fig:CS_tot_NOambiguities} (since we fixed the amplitudes to those for the $++$ contribution) but one does not expect a direct delta exchange here. When the delta contribution is removed from $J_{\pm,\mp}$, the relative magnitude of $\lvert J_{\pm,\mp}\rvert$ to $\lvert J_{++} \rvert$ reduces to percent level.\!\footnote{We checked explicitly but crudely by setting the multipole amplitudes for quantum numbers of the $\Delta$ resonance to zero.}

\section{Conclusions \label{sec:conclusions} } 
Relativistic mean field models are useful tools for modeling reactions with nuclei, and relating bound-state properties to those that can be measured on free nucleons. Often one encounters difficulties when using the standard prescription, \cref{prescription},  because Gordon-like identities which leave the free-nucleon matrix elements unchanged can alter the bound-state matrix elements. Our main conclusion is that these so-called ``off-shell ambiguities'' are spurious consequences of the often used ansatz given in \cref{prescription}.

A Bogoliubov transformation relates bound-state creation operators to free particle (anti-)nucleon creation and annihilation operators. This yields a relation between free and bound-state matrix elements free of ambiguities related to on-shell algebra. This does not address the dependence of the current on the mean-field potentials themselves, nor on their operator level definition (which is often not uniquely fixed by on-shell matrix elements).

With this assumption, we  have presented formulae which relate the bound-state matrix elements to explicitly on-shell operator structures; 
Our main result is \cref{main-result} which provides a mapping between free-nucleon and bound-state matrix elements. This result is manifestly free from ambiguities related to on-shell Dirac algebra. One sees explicitly that the presence of projectors guarantees that all identities that use the free Dirac equation (e.g. the Gordon identity) leave the resulting amplitudes invariant. 

We have discussed and derived some results that follow directly from \cref{main-result}. 
First, one obtains \cref{prescription} from \cref{main-result} when the condition of \cref{eq:condition} is met e.g., for elementary currents as discussed in \cref{sec:elementary_my_dear}.
Second, there are contributions from matrix elements for creation/annihilation of $N\bar{N}$ pairs due to the non-trivial mapping between the free vacuum and the vacuum in the presence of mean-field potentials cf. \cref{eq:interacting_vacuum}.
Third, one may define positive/negative energy projections of bound state wave functions based on \cref{main-result}, given explicitly in \cref{app:useful-formulae}.  One may identify the leading $++$ contribution, i.e. the $\Gamma^{(a)}$ term in \cref{main-result}, with the ``typical'' or ``textbook'' non relativistic treatment  \cite{walecka04} upon expansion in powers $|\vb{p}|/M$, as shown in \cref{sec:Currents_w_FF}.

We have also discussed possible ramifications for phenomenology for coherent interactions, elastic electron scattering and pion production.

For elastic scattering, we studied the full nuclear current and its treatment when using on-shell Gordon-like identities. In general, when using an arbitrary representation of the current, \cref{prescription} fails. If, however, a representation is chosen such that there is no explicit momentum dependence {\it and} nucleon form factors are evaluated at a single fixed momentum transfer then \cref{prescription} agrees with \cref{main-result}. 
The nuclear charge form factor for a closed shell is then given in terms of the Fourier transforms of the vector and tensor densities.
The dominant contribution is obtained by retaining only positive energy states. When it is expanded up to $O(\vb{p}^2/M^2)$ it may be written in terms of standard multipole operators~\cite{walecka04, Waleckapaper} yielding the $\mathcal{F}_{M/\Phi^{\prime\prime}}$ responses of Ref.~\cite{Hoferichter:2020osn}. We show that both the momentum dependence of nucleon form factors around the central value $\lvert\vb{q}\rvert$ and the purely negative energy ($--$) contribution yield corrections of order $p^4/M^4$.
We find a consequential difference between the use of \cref{prescription} and \cref{main-result} which stems from the positive-negative ($\pm,\mp$) contributions. When using \cref{prescription}, these terms are large enough to noticeably shift the position of the first zero of the nuclear form factor obtained by only including positive energy states. 
The Bogoliubov transformation that relates the free creation and annihilation operators to those derived in the presence of background fields, demands that the form factors in the $\pm,\mp$ terms should be evaluated at time-like four-momenta. We consider 7 different parametrizations of the time-like form factors and find that in each case the $\pm,\mp$ terms are strongly suppressed. The final result is essentially indistinguishable from retaining only positive energy states which differs from what one would obtain using \cref{prescription}.

For coherent pion photoproduction we studied the leading contribution obtained by retaining only positive energy states. We provide a fully relativistic treatment, including spin flip terms for closed-shell nuclei. The latter are not coherently enhanced and are found to be negligible for calcium, but non-negligible for carbon. The leading non spin-flip term is dominated by the $F_2$ CGLN amplitude. We show that retaining only this term is an excellent approximation. Large discrepancies with the complete result are found when the single-nucleon amplitude is not averaged over the nucleon momentum.
We confirm the $O(1)$ changes in the cross section pointed out in Ref.~\cite{Abu-Raddad:1998ams} when using \cref{prescription} with different arbitrary choices of the basis of invariant amplitudes used to parametrize free nucleon matrix elements. These changes stem from the $(\pm,\mp)$ terms.
Using \cref{main-result} we obtain results that are both derived from the operator definition of the current and {\it manifestly independent} of the equivalent parametrizations of the on-shell matrix elements. 
When the $(\pm,\mp)$ terms are evaluated using \cref{main-result} but with the invariant amplitudes fixed to those for direct pion production ($\gamma N \rightarrow \pi N$), one obtains unique but sizable contributions. They are $O(10\%)$ of the leading $++$ \emph{amplitude} as shown in \cref{fig:CS_tot_NOambiguities}.
We argue that these should be expected to be reduced by an order of magnitude when evaluated at timelike four-momentum transfer.\!\footnote{The same procedure in elastic scattering, i.e. including invariant amplitudes (form factors) at space-like momentum transfer, overestimates the $\pm,\mp$ contributions by an order of magnitude compared to the result obtained when using experimentally determined time-like form factors.}

Retaining the leading order positive energy contribution therefore provides a drastic reduction in the uncertainty for coherent pion photoproduction as compared to the use of the {\it ansatz} in  \cref{prescription}.
Our approach allows to compute relativistic corrections arising from the negative energy states in a systematic way. These are expected to be sub-leading to effects such as in-medium modification of resonance properties, final-state interactions, and higher order spin-flip and charge-exchange contributions~\cite{Tsaran:2024sue, Miller:2019btv, 10.1143/ptp/86.6.1277, TIATOR1980343, CARRASCO1993797}.
This is important for the modeling of electromagnetic and neutrino induced pion production on nuclei with relativistic mean field models~\cite{Nikolakopoulos:2022tut, Gonzalez-Jimenez19, Garcia-Marcos:2023rnj}.

The results of \cref{sec:elastic} may be important in the context of parity violating electron scattering. We find that relativistic mean-field models cannot automatically use \cref{prescription}; subleading contributions from time-like form factors affect the results. This is important when attempting an extrapolation of neutron skin measurements in medium and heavy nuclei to neutron stars (which are in effect nuclear matter) \cite{Mammei:2023kdf}. This extrapolation relies on nuclear mean-field modeling and it would be interesting to apply the approach outlined here to that problem.

In summary, we have investigated a well known source of uncertainty in mean-field models. We find that when currents are defined at an operator level and the relation between free-particle states and bound states is carefully worked out, that all ``off-shell ambiguities'' disappear. Other ambiguities related to the current's dependence on the background fields and the correct operator level definition of the current are not addressed. These results are important for phenomenological applications of mean field models. 

\textbf{Acknowledgments:} R.P. thanks Mark Wise for many useful and illuminating discussions. A.N. thanks Raul Gonzalez-Jimenez, Natalie Jachowicz and Matthias Hooft for helpful discussions. A.N. is supported by the
Neutrino Theory Network (NTN) under Award Number DE-AC02-07CH11359. 

\onecolumngrid

\appendix
\crefalias{section}{appendix}
\section{Coherent interactions \label{app:useful-formulae} }
We derive here useful formulae for the calculation of \cref{main-result} for coherent interactions.
The bound-state wavefunctions in momentum space have the form
\begin{equation}\label{eq:upper-lower}
    \psi_\alpha(\vb{p}) = \begin{pmatrix}
        g_{n,\kappa}(\lvert \vb{p} \rvert) \Phi_\kappa^{m}(\Omega_{p}) \\
        f_{n,\kappa}(\lvert \vb{p} \rvert) \vb*{\sigma}\cdot\hat{\vb{p}}\Phi_\kappa^{m}(\Omega_{p})
    \end{pmatrix}
\end{equation}
where the quantum numbers $\alpha = \{ n, \kappa, m \}$, are the principal quantum number, the angular momenta $\kappa$, and the angular momentum projection $m$. 
The total angular momentum is $j =\lvert \kappa \rvert - 1/2$, and orbital angular momentum is $l = j \pm 1/2$ for $\kappa = \pm \lvert \kappa \rvert$. The spin-spherical harmonics $\Phi_\kappa^m$ are defined in~\cite{Bechler_1993}.
We suppress the subscript $n,\kappa$ and argument $\lvert \vb{p} \rvert$ for the radial wavefunctions in the following for brevity.
The positive/negative energy projections are readily computed and matrix elements may be conveniently  written in terms of states
\begin{equation}
    \psi_{\alpha,\pm} \equiv \tilde{g}_{\pm}(\lvert \vb{p} \rvert) u_{\pm}(\vb{p}) \Phi_\kappa^{m}(\Omega_p),
\end{equation}
were we have defined radial wavefunctions
\begin{equation}
\label{eq:def_radial}
\tilde{g}_\pm(p) = \frac{(E\pm M)}{2E}g \pm \frac{p}{2E}f,
\end{equation}
and the positive/negative energy spinor projectors
\begin{equation}
\label{eq:def_uplusmin}
    u_{\pm}(\vb{p}) = \begin{pmatrix}
1 \\
\pm \frac{\vb*{\sigma}\cdot\vb{p}}{E\pm M}
    \end{pmatrix}
\end{equation}
with $E = \sqrt{p^2 + M^2}$.
These states are defined such that
\begin{equation}
    \psi_{\alpha,+} = \frac{(\slashed{p} + M)}{2E}\gamma^0 \psi_\alpha, \quad \overline{\psi}_{\alpha,+} = \psi_\alpha^\dagger\frac{(\slashed{p} + M)}{2E},
\end{equation}
\begin{equation}
    \psi_{\alpha,-} = \frac{(\tilde{\slashed{p}} - M)}{2E}\gamma^0 \psi_\alpha, \quad \overline{\psi}_{\alpha,-} = \psi_\alpha^\dagger\frac{(\tilde{\slashed{p}} - M)}{2E}.
\end{equation}
One sees that the appearance of the $\gamma^0$ in \cref{main-result} is quite natural, without it one would have a sign change in the $f$ contribution in \cref{eq:def_radial} depending on whether the state enters on the left or right of the bilinear.

Let us now consider coherent interactions, in this case $\alpha = \beta = \{n,\kappa,m\}$, and a coherent sum over the states is performed
To perform the sum over angular momenta, it is convenient to define the two-by-two matrices $\mathcal{F}_{s,s^\prime}$ as
\begin{equation}
\label{eq:defmathFPauli}
     \chi^{\prime} \mathcal{F}_{s,s^\prime}(\vb{p}^\prime,\vb{p}) \chi \equiv \chi^{\prime} \overline{u}_{s^\prime}(\vb{p}^\prime) \Gamma(\vb{p},\vb{p}^\prime) u_{s}(\vb{p}) \chi
\end{equation}
where $s,s^\prime \in \{+,-\}$ denote positive/negative energy projections of initial and final-state wavefunctions respectively. 
Here $\chi^{(\prime)}$ are two-component Pauli spinors, and $u_{s}$ are given in \cref{eq:def_uplusmin}.
One can then write the contributions to the current as
\begin{equation}
\label{eq:current_coherent}
    J_{s,s^\prime} = \sum_{n,\kappa,m} \langle n, \kappa,m  \rvert \hat{J}_{ss^\prime} \lvert n, \kappa, m \rangle = \sum_{n,\kappa} \int \frac{\mathrm{d}\vb{p}^\prime \mathrm{d}\vb{p}}{(2\pi)^6} \tilde{g}_{n,\kappa,s^\prime}(\lvert \vb{p}^\prime \rvert) \tilde{g}_{n,\kappa,s}(\lvert \vb{p} \rvert) \mathrm{Tr}\left[ \mathcal{F}_{s,s^\prime}(\vb{p}^\prime, \vb{p}) \left(\sum_{m} \Phi_\kappa^{m}(\Omega_p) \Phi_{\kappa}^{m,\dagger}(\Omega_{p^\prime}) \right) \right].
\end{equation}
The sum over spinor spherical harmonics can be readily performed~\cite{Bechler_1993} yielding
\begin{equation}
\label{eq:sum_Phikappa}
 \left(\sum_m \Phi_\kappa^{m} \Phi_\kappa^{m,\dagger} \right) = \frac{1}{4\pi}  \left( (j+1/2) P_l(\hat{\vb{p}}\cdot\hat{\vb{p}}^\prime) - i \frac{\kappa}{\lvert\kappa \rvert} P_{l}^{\prime} (\hat{\vb{p}}\cdot\hat{\vb{p}}^\prime)~\vb*{\sigma}\cdot\left(\hat{\vb{p}}\times\hat{\vb{p}}^\prime \right) \right),
\end{equation}
where $P_l$ are the Legendre polynomials and $P_l^\prime$ denotes the derivative.
Writing
\begin{equation}
    \mathcal{F}_{s,s^\prime} = F_{s,s^\prime}(\vb{p},\vb{p}^\prime) + i\vb*{\sigma}\cdot\vb{K}_{s,s^\prime}(\vb{p}^\prime, \vb{p}),
\end{equation}
the traces in \cref{eq:current_coherent} are readily performed and one identifies a coherently enhanced contribution, and a contribution from spin-orbit effects.

\section{Elastic Scattering}
\label{app:elastic_scattering}
We consider elastic scattering, for which the Dirac bilinears are given in \cref{eq:gammaa_elastic,eq:gammac_elastic}. 
Since they all have the same structure, we can consider the same bilinear
\begin{equation}
    \Gamma^\mu = g_1 \gamma^\mu + \iu \frac{g_2}{2M} \sigma^{\mu\nu}Q_\nu~,
\end{equation}
for each case. The difference between the $\pm,\pm$ and $\pm,\mp$ contributions are the scalar functions $g_1$ and $g_2$. We write $Q^\mu = (0,\vb{q})$.
For $\Gamma^0$ we have
\begin{equation}
    \mathcal{F}^0_{s,s^\prime}(\vb{p}^\prime,\vb{p}) = g_1\left[1 +  \frac{\vb*{\sigma}\cdot\vb{p}^\prime \vb*{\sigma}\cdot \vb{p}}{N_s N_{s^\prime}}\right] + \frac{g_2}{2M} \left[ \frac{\vb*{\sigma}\cdot{\vb{q}} \vb*{\sigma}\cdot \vb{p}}{N_{s}} -   \frac{\vb*{\sigma}\cdot{\vb{p}^\prime} \vb*{\sigma}\cdot \vb{q}}{N_{s^\prime}}\right]~,
\end{equation}
where we defined $N_{\pm^{({\prime})}} = \pm (E^{(\prime)}\pm M)$.
Using the result in \cref{eq:sum_Phikappa}, the traces in \cref{eq:current_coherent} can be readily performed.
Adding the momentum conserving delta function one then has
\begin{equation}
\label{eq:J0ssprime_result}
    J_{s,s^\prime}^0 = \sum_{\kappa,m} \langle \kappa ,m \rvert \hat{J}^0_{s,s^\prime} (\lvert \vb{q} \rvert) \lvert \kappa, m \rangle =  \sum_{\kappa} \int \frac{\mathrm{d}\vb{p}}{(2\pi)^3} \langle \kappa \rvert \mathcal{J}_{s,s^\prime}^0(\vb{p} + \vb{q} ,\vb{p}) \lvert \kappa \rangle,
\end{equation}
with the current densities for each $\kappa$ given by
\begin{align}
    \langle \kappa \rvert \mathcal{J}_{s,s^\prime}^0(\vb{p}^\prime ,\vb{p}) \lvert \kappa \rangle &= \tilde{g}_{s^\prime}(\lvert \vb{p}^\prime \rvert ) \tilde{g}_{s}(\lvert \vb{p} \rvert ) \frac{2J+1}{4\pi} P_l(\hat{\vb{p}}\cdot\hat{\vb{p}}^\prime)\left[ g_1 ( 1 + \frac{\vb{p}\cdot\vb{p}^\prime}{N_s N_{s^\prime}})  + \frac{g_2}{2M}\vb{q} \cdot (\frac{\vb{p}}{N_s} - \frac{\vb{p}^\prime}{N_{s^\prime}})  \right] \nonumber \\
    &+ \frac{\kappa}{\lvert \kappa \rvert}\frac{\tilde{g}_{s^\prime}(\lvert \vb{p}^\prime \rvert ) \tilde{g}_{s}(\lvert \vb{p} \rvert )}{4\pi} P_l^\prime(\hat{\vb{p}}\cdot\hat{\vb{p}}^\prime)\frac{\lvert \vb{p}^\prime \cross \vb{p} \rvert^2}{\lvert \vb{p} ^\prime \rvert \lvert \vb{p} \rvert}\left[ \frac{g_1}{N_sN_{s^\prime}}  +  \frac{g_2}{2M}\left ( \frac{1}{N_s} + \frac{1}{N_{s^\prime}} \right)  \right]~.
    \label{eq:Jdens_result}
\end{align}
The full momentum dependence of the form factors in \cref{eq:gammaa_elastic,eq:gammac_elastic}, can be readily included by substituting $g_i \rightarrow F_i(\vb{q}^2 - (E - E^\prime)^2)$ in the $\pm,\pm$ contributions and $g_i \rightarrow F_i(\vb{q}^2 - (E + E^\prime)^2)$ in the $\pm,\mp$ contributions.

Let us briefly examine the spatial current $\vb{e}\cdot\vb{J}$, which should disappear when recoil is neglected. 
We use shorthand
\begin{equation}
    \vb{e} \cdot \vb{J}_{s,s^\prime}(\vb{q}) = \int_{\vb{p}} \mathcal{J}_{s,s^\prime} (\vb{p} + \vb{q}, \vb{p})~,
\end{equation}
where the current density $\mathcal{J}_{s,s^\prime}(\vb{p}^\prime, \vb{p})$ is defined through Eq.~(\ref{eq:current_coherent}) in the following.
It is straightforward\footnote{For example: each of the current densities is a scalar which can be written $\vb{e}\cdot\langle \kappa \rvert \vb{\mathcal{J}}_{s,s^\prime}(\vb{p} + \vb{q} ,\vb{p}) \lvert \kappa \rangle = A(\vb{p},\vb{q}) \vb{e}\cdot\vb{q} + B(\vb{p},\vb{q}) \vb{e}\cdot\vb{p}$, with $A,B$ scalar functions constructed from $\vb{q}, \vb{p}$. In a spherical coordinate system with $\vb{q}$ defining the $z$-axis, i.e. $\vb{q}\cdot\vb{p} = \lvert \vb{p} \rvert \lvert \vb{q}\rvert \cos\theta_p$, we have $\vb{e}\cdot\vb{p} \propto a\sin\phi + b\cos\phi$, which disappear after integrating over the azimuthal angle.} to see that when $\vb{e}\cdot \vb{q} = 0$, $\vb{e}\cdot \vb{J}_{s,s^\prime} = 0$ for each combination of $s,s^\prime$.
Hence we only need to consider $\vb{e} = \hat{\vb{q}}$. The $g_2$ term is trivially zero when recoil is neglected (proportional to $\omega =0$), so we only examine the $g_1$ contribution.
Inserting
\begin{equation}
    \vb{e}_i\vb{\mathcal{F}}^i_{s,s^\prime} = g_1 \left[  \frac{\vb*{\sigma}\cdot\vb{p}^\prime \vb*{\sigma}\cdot\vb{e}}{N_{s^\prime}} + \frac{\vb*{\sigma}\cdot\vb{e}\vb*{\sigma}\cdot\vb{p}}{N_s} \right],
\end{equation}
in Eq.~(\ref{eq:current_coherent}) one sees that,
\begin{equation}
\label{eq:symmetry_current_TR}
    \mathcal{J}_{+-}(\vb{p}^\prime,\vb{p}) = \mathcal{J}_{-+}(\vb{p},\vb{p}^\prime).
\end{equation}
This is the case specifically for coherent interactions, since the combination of radial wavefunctions $\tilde{g}_{\pm}(\vb{p}^\prime) \tilde{g}_{\mp}(\vb{p})$ that enters in Eq.~(\ref{eq:current_coherent}) satisfies this property.
We can then see that
\begin{equation}
    \vb{e}\cdot\vb{J}_{+-}(\vb{q}) = \int_{\vb{p}} \mathcal{J}_{+-}(\vb{p+\vb{q}}, \vb{p}) = \int_{\vb{p}} \mathcal{J}_{+-}(\vb{p}, \vb{p}-\vb{q}) = \int_{\vb{p}} \mathcal{J}_{-+}(\vb{p}-\vb{q}, \vb{p}) = \vb{e}\cdot\vb{J}_{-+}(-\vb{q}) = -\vb{e}\cdot\vb{J}_{-+}(\vb{q}).
\end{equation}
The second equality is substitution, the third uses the property of Eq.~(\ref{eq:symmetry_current_TR}), and the last follows from the fact that the current transforms as a vector under parity. Eq.~(\ref{eq:longitudinal_cancellation_elastic}) follows taking $\vb{e} = \vb{q}$ and the current is conserved as required.
A similar argument applied to the time-like part gives $J_{-+}^0 = J_{+-}^0$, as can be seen from Eq.~(\ref{eq:Jdens_result}).

\section{Coherent pion production}
\label{app:cohpi}
Results for pion production are obtained by taking $\vb*{\epsilon} \cdot \vb*{\mathcal{F}}_{ss^\prime}$ as defined in \cref{eq:defmathFPauli} and computing the traces and integrals of \cref{eq:current_coherent} with a delta function $(2\pi)^3\delta(\vb{p}^\prime - \vb{p} + \vb{k}_\pi - \vb{q})$. 
Using the invariant amplitudes associated with the matrices in Eqs.~(\ref{eq:M1_def}-\ref{eq:M4_def}) one finds 
\begin{align}
    \vb*{\epsilon}\cdot \vb*{\mathcal{F}}_{ss^\prime}(\vb{p},\vb{q},\vb{p}^\prime) &= - \frac{A_1}{N_{s^\prime}} \vb*{\epsilon}\cdot(\vb{q}\cross \vb{t}) + A_1 \left(\frac{1}{N_{s}} - \frac{1}{N_{s^\prime}} \right) \vb*{\epsilon}\cdot (\vb{q}\cross\vb{p}) +  \frac{q^0 A_1}{N_{s}N_{s^\prime}} \vb*{\epsilon}\cdot (\vb{p}\cross \vb{t}) + \frac{G_{3}}{N_{s}N_{s^\prime}} \vb*{\epsilon}\cdot(\vb{p}\cross\vb{t}) + \frac{G_4}{N_{s}N_{s^\prime}} \vb{q}\cdot(\vb{p}\cross\vb{t}) \nonumber \\
    &- \iu\vb*{\sigma}\cdot\vb*{\epsilon} \left[  G_3 - A_1\left( q^0 - \left(\frac{1}{N_{s}} + \frac{1}{N_{s^\prime}}\right)\vb{q}\cdot\vb{p} - \frac{\vb{q}\cdot\vb{t}}{N_{s^\prime}} \right) - \frac{q^0 A_1 + G_3}{N_{s}N_{s^\prime}} \left( \vb{p}^2 + \vb{p}\cdot\vb{t} \right)\right] \nonumber \\
    &+ \iu\vb*{\sigma}\cdot\vb{q} \left[ A_1 \left( \vb*{\epsilon}\cdot\vb{p} \left( \frac{1}{N_s} + \frac{1}{N_{s^\prime}}\right)  + \frac{\vb*{\epsilon}\cdot\vb{t}}{N_{s^\prime}} \right) + G_4\left( \frac{\vb{p}^2 + \vb{p}\cdot\vb{t}}{N_{s}N_{s^\prime}} - 1 \right)\right] + \vb*{\sigma}\cdot\vb{p}\left[\ldots \right] + \vb*{\sigma}\cdot \vb{t} \left[ \ldots \right]~,
    \label{eq:edotJ_Aamps}
\end{align}
where again $N_{\pm^{(\prime)}} = \pm \left( E^{(\prime)} \pm M \right)$ and the $[\ldots]$ in \cref{eq:edotJ_Aamps} can be inferred from Eq.~(35b) of Ref.~\cite{BENNHOLD1991625}.
Equivalent results have been given previously in Ref.~\cite{BENNHOLD1991625}, and we have defined the shorthand 
\begin{equation}
    G_3 \equiv (A_3 + A_4)(q^0E - \vb{q}\cdot\vb{p}) + A_4\left[ (E^\prime-E)q^0 - \vb{p}\cdot\vb{t}\right],\quad G_4 \equiv (A_3 + A_4)\vb*{\epsilon}\cdot\vb{p} + A_4 \vb*{\epsilon}\cdot\vb{t}
\end{equation}
following that reference. 
The only difference is that we have  written $\vb{t} = \vb{p}^\prime - \vb{p} = \vb{q} - \vb{k}_\pi$, and the explicit appearance of $q^0$.
The latter follows from the on-shell treatment where $q^0 = E_\pi + E^\prime - E \neq \lvert \vb{q} \rvert$.
The terms proportional to $\vb*{\sigma}\cdot\vb{p}$ and $\vb*{\sigma}\cdot\vb{t}$ are not given explicitly since these  do not contribute to the coherent interaction\footnote{
The traces in \cref{eq:current_coherent} for these terms are proportional to the triple product $\vb{p}\cdot(\vb{p}\cross\vb{t})$ and hence disappear identically.}.

Inserting this form into Eq.~(\ref{eq:current_coherent}) yields a current consistent with Eq.~(\ref{eq:generalJmu}) since all terms that are sensitive to components of $\vb*{\epsilon}$ orthogonal to $\vb{q}\cross\vb{k}_\pi$ (e.g. $\vb*{\epsilon}\cdot(\vb{q}\cross \vb{p})$) involve integrals over odd functions of the nucleon azimuth angle that disappear. As a result one can simplify further by dropping the $\vb*{\epsilon}\cdot\vb{t}$ terms.

\subsection{Relation to CGLN amplitudes}
\label{app:CGLN}
The invariant amplitudes $A_i$ for pion photoproduction can be related to the CGLN amplitudes~\cite{Chew:1957tf}.
The latter parametrize matrix elements computed in the CMS, where $\vb{q}^*=-\vb{p}^*$.
They are defined as
\begin{equation}
    \iu \epsilon\cdot\mathcal{F}_{\rm CGLN} = F_1~\vb*{\sigma}\cdot\vb*{\epsilon}  -\iu F_2~(\vb*{\sigma}\cdot \hat{\vb{k}}_\pi^*) \vb*{\sigma}\cdot(\hat{\vb{q}}^* \cross \vb*{\epsilon}) + F_3~(\vb*{\sigma}\cdot \hat{\vb{q}}^*) \hat{\vb{k}}_\pi\cdot\vb*{\epsilon} + F_4~(\vb*{\sigma}\cdot \hat{\vb{k}}_\pi^*) \hat{\vb{k}}_\pi^*\cdot\vb*{\epsilon}.
    \label{eq:F_CGLNdef}
\end{equation}
The normalization is given by 
\begin{equation}
    \frac{\mathrm{d}\sigma}{\mathrm{d}\Omega^*} = \frac{\lvert \vb{k}_\pi^* \rvert}{\lvert \vb{q}^*\rvert} \frac{1}{4} \sum_{m,m^\prime,\lambda} \lvert \chi_{m^\prime}^\dagger~ \epsilon_\lambda \cdot \mathcal{F}_{\rm CGLN}~\chi_m \rvert^2
\end{equation}

The relation to the amplitudes in \cref{eq:edotJ_Aamps} is obtained by evaluating the latter in the CMS for on-shell photons $q^0=\lvert \vb{q} \rvert$.
The $A_i$ amplitudes are normalized in the conventional manner~\cite{Bjorken}, such that
\begin{equation}
    \frac{\sqrt{(E^* + M)(E^{\prime *} + M)}}{8\pi W} \vb*{\epsilon} \cdot \vb*{\mathcal{F}}_{++}(-\vb{q}^*,\vb{q}^*, -\vb{k}_\pi^*) = \epsilon\cdot\mathcal{F}_{\rm CGLN},
\end{equation}
where $W^2 = (p + q)^2 = (p^\prime + k_\pi)^2.$
One obtains the relations given in Ref.~\cite{BENNHOLD1991625}.
We give only the expression for the leading term of \cref{eq:F0_leading} for later discussion
\begin{equation}
\label{eq:A1_cgln}
A_1 = \frac{4\pi}{\lvert\vb{q}^*\rvert\sqrt{2W}} \left[ F_1 \frac{W-M}{\sqrt{E^{\prime*} + M_N}} - F_2\frac{W+M}{\sqrt{E^{\prime*} - M_N}}\right].
\end{equation}
The CGLN amplitudes may be written in terms of magnetic and electric multipole amplitudes~\cite{Chew:1957tf}.
We use the multipole amplitudes obtained from the ANL-Osaka DCC model~\cite{DCC:electron, Kamano:2019gtm} to construct the invariant amplitudes $A_i$.

We may avoid this 'intermediate step' and write $\vb*{\epsilon}\cdot \vb*{\mathcal{F}}_{++}$ in the lab frame directly in terms of CGLN amplitudes.
The boost that connects the lab-frame to the CMS is the inverse boost with velocity
\begin{equation}
\vb*{\beta} = \frac{\vb{q} + \vb{p}}{q^0 + E} = \frac{\vb{p}^\prime + \vb{k}_\pi}{E^\prime + E_\pi}.
\end{equation}
The CMS momenta are obtained from lab-frame momenta as
\begin{equation}
    \vb{k}^* =  \vb{k} + \hat{\vb*{\beta}}\left[ \hat{\vb*{\beta}}\cdot\vb{k} (\gamma - 1)  - k^0 \lvert \vb*{\beta} \rvert \gamma \right] = \vb{k} + \vb*{\beta} \tilde{B}(k),
\end{equation}
where $\gamma = (1-\vb{\beta}^2)^{-1/2}$.
Note that the on-shell treatment means that there is no ambiguity on what boost needs to be applied~\cite{PhysRevC.21.1472}.
Finally, a Wigner rotation should be applied to the spin states, described in detail in \cref{sec:Wigner_rotation}.
The current may then be evaluated at CMS momenta and  boosted back to the lab frame~\cite{Rocco:2019gfb}.

Instead it is more straightforward to boost the polarization vector $\epsilon$ to the CMS.
Under a boost with velocity $\vb*{\beta}$ the polarization vector transforms by the rotation~\cite{PhysRevD.31.328, PhysRev.135.B1049},
\begin{equation}
    \vb*{\epsilon}^* = \vb*{\epsilon} - \frac{\vb*{\epsilon}\cdot\vb*{\beta}}{1+\vb*{\beta}\cdot\hat{\vb{q}}}\left( \hat{\vb{q}} + \frac{\gamma}{1+\gamma} \vb*{\beta} \right).
\end{equation}
This is the rotation that takes the direction of the momentum of a real photon in the lab frame $\hat{\vb{q}}$ to the direction in the CMS $\hat{\vb{q}}^*$ applied specifically to a vector $\vb*{\epsilon}$ that satisfies $\vb*{\epsilon}\cdot\vb{q}=0$.
In summary the matrix elements are given by
\begin{equation}
   \mathcal{N}~\vb*{\epsilon}\cdot \vb*{\mathcal{F}}_{++}(\vb{p}, \vb{q},\vb{p}^\prime) =  [R(\theta^\prime)]^\dagger \vb*{\epsilon}^*\cdot \vb*{\mathcal{F}}_{\rm CGLN}(\vb{q}^*, \vb{k}_\pi^*) R(\theta),
\end{equation}
with $\mathcal{N}$ some conventional normalization, and $R(\theta^{(\prime)})$ are the Wigner rotations of the initial(final) nucleon spins given explicitly in \cref{sec:Wigner_rotation}.

This expression can be readily evaluated numerically, but it is useful to consider some limiting cases.
All the boosts introduce corrections of order $(\vb{q} + \vb{p})/(q^0 + M)$ or smaller, including the Wigner rotations.
If one neglects them completely, it is clear that the coherently enhanced contribution (the trace of $\vb*{\epsilon}\cdot\vb*{\mathcal{F}}_{++}$) is given solely in terms of $F_{2}~\vb*{\epsilon} \cdot (\vb{q} \cross \vb{t})$, all other contributions are traceless.

The 'leading' contribution in terms of $A_i$ given in Eq.~(\ref{eq:F0_leading}) may be identified as the trace of \cref{eq:edotJ_Aamps} for $\vb{p}=\vb{0}$.
It is given solely in terms of $A_1$, which contains $F_2$ in addition to a suppressed contribution from the $F_1$ CGLN amplitude as seen in \cref{eq:A1_cgln}.
The boosts of vectors do not affect the trace, only the spin-rotation does.
For $\vb{p}=\vb{0}$ there is no rotation for the initial nucleon spin. The rotation for the final nucleon will introduce a term proportional to $\sigma\cdot (\vb{q}\cross\vb{t})$ which then gives a small (cf. \cref{eq:Wignersmall} ) $F_1$ contribution consistent with \cref{eq:A1_cgln}. 

\subsection{Wigner rotation}
\label{sec:Wigner_rotation}
To connect the (positive energy) wavefunctions to the CGLN amplitudes one performs a boost to the pion-nucleon CMS system; this implies a spin rotation.
Indeed, with $B^{(-)}(p)$ denoting an (inverse) boost with four-momentum $p$ i.e. velocity $\vb{p}/p^0$, the boost of the positive energy wavefunctions to the CMS may be written
\begin{equation}
     B^{-}(p+q) u_+(\vb{p}) \Phi = B^{-}(p+q) B(p) u_+(0)\Phi = B(p^*) R(\theta) u_+(0) \Phi = u_+(\vb{p}^*) R(\theta)\Phi = u_{+}(\vb{p}^*)\Phi_{\rm cm}~,
\end{equation}
since two boosts can be decomposed into a single boost and a rotation. The positive energy spinor $u_+(\vb{p}^*)$ is included in the CGLN decomposition, we only need to include the spin rotation.
The Wigner rotation acting on two-component spin states obtained from two subsequent boosts with four-momenta $p,P$ respectively is~\cite{Polyzou:2014yea}
\begin{equation}
    R(\theta) = \cos\frac{\theta}{2} + \iu \vb*{\sigma}\cdot\hat{\vb{n}}\sin\frac{\theta}{2} = B^{-}(p^*) B(P) B(p)~,
\end{equation}
Where $p^*$ is given by the Lorentz boost $p^{*\mu}=\Lambda_{~\nu}^\mu[B(P)]p^\nu$.
The boosts are given by
\begin{equation}
    B(p) = \cosh \frac{\xi}{2} + \sinh \frac{\xi}{2} \vb*{\sigma}\cdot\hat{\vb*{\beta}}~, \quad B^-(p) = \cosh \frac{\xi}{2} - \sinh \frac{\xi}{2} \vb*{\sigma}\cdot\hat{\vb*{\beta}}~,
\end{equation}
where $\vb*{\beta} = \vb{p}/p^0$, $\cosh\xi = \gamma = 1/\sqrt{1-\vb*{\beta}^2} = p^0/\sqrt{p^2}$, $\sinh \xi = \lvert \vb*{\beta} \rvert \gamma = \lvert \vb{p} \rvert/ p^0$, and thus
\begin{equation}
    \tanh \frac{\xi}{2} = \frac{\sinh\xi}{\cosh\xi + 1} = \frac{\sqrt{p^2}}{p^0} \frac{\lvert \vb{p} \rvert}{p^0 + \sqrt{p^2}}~.
\end{equation}
By inspection of two consecutive boosts, with velocities $\vb*{\beta}$, $\vb*{\beta}^\prime$ respectively one readily finds that the rotation axis and angle are explicitly given by
\begin{equation}
    \hat{\vb{n}} = \frac{\hat{\vb*{\beta^\prime}}\cross\hat{\vb*{\beta}}}{\sqrt{1-(\hat{\vb*{\beta}}\cdot\hat{\vb*{\beta^\prime}})^2}}~, \quad \tan\frac{\theta}{2} = \frac{\tanh\frac{\xi}{2} \tanh\frac{\xi^\prime}{2}\sqrt{1-(\hat{\vb*{\beta}}\cdot\hat{\vb*\beta^\prime})^2}  }{1 + \tanh\frac{\xi}{2} \tanh\frac{\xi^\prime}{2}\hat{\vb*{\beta}}\cdot\hat{\vb*{\beta^\prime}}}~.
\end{equation}
Applying this now to the boost to the CMS for pion production with the initial(final) nucleon momentum in the lab frame denoted as $\vb{p}^{(\prime)}$,
\begin{equation}
    \tanh\frac{\xi^\prime}{2}\tanh \frac{\xi}{2} = \left[\frac{W_{\pi N}}{E^\prime + E_\pi} \frac{\lvert \vb{p} + \vb{q} \rvert}{E_\pi + E^\prime + W_{\pi N}}\right]\frac{\lvert\vb{p}^{(\prime)} \rvert}{E^{(\prime)} + W_{\pi N}}~.
\end{equation}
Such that the magnitude of the rotation is given by
\begin{equation}
\label{eq:Wignersmall}
    \sin \frac{\theta}{2} = \frac{\tan\frac{\theta}{2}}{\sqrt{1 + \tan^2\frac{\theta}{2}}} \simeq \left[\frac{W_{\pi N}}{E^\prime + E_\pi} \frac{\lvert \vb{p} + \vb{q} \rvert}{E_\pi + E^\prime + W_{\pi N}}\right]\frac{\lvert\vb{p}^{(\prime)} \rvert}{E^{(\prime)} + W_{\pi N}} + O\qty( \frac{\vb{p}^4}{M^4})~,
\end{equation}
and
\begin{equation}
    \cos \frac{\theta}{2} = \frac{1}{\sqrt{1 + \tan^2\frac{\theta}{2}}} \simeq 1 + O\qty( \frac{\vb{p}^4}{M^4}).
\end{equation}

\section{Time-like form factors and Compton Scattering \label{app:compton-example}}
The appearance of time-like form factors in the bound-state matrix elements may appear surprising. In this section we describe how the same objects naturally appear when considering Compton scattering in time-ordered perturbation theory, and are therefore not ``peculiar'' to relativistic bound states. Since we work with explicit time orderings, energy is not conserved at each vertex and all intermediate states are on-shell. 

Let us consider Compton scattering on a proton, 
\begin{equation}
    \gamma(\vb{k}) + p(\vb{0}) \rightarrow \gamma(\vb{k}') + p(\vb{Q})~, 
\end{equation}
with $\vb{k}'= \vb{k}-\vb{Q}$. The Compton tensor is typically defined by (we assume a possible seagull term is absent e.g., with currents defined in terms of elementary quarks)
\begin{equation}
    H_{\mu\nu} = \int \dd^4x ~\e^{-\iu k\cdot x} \mel{p(\vb{Q})}{T\{ J_{\mu}(x) J_\nu(0)\}}{p(\vb{0})}~. 
\end{equation}
Writing time-orderings explicitly, translating all operators to $x=0$, and inserting a complete set of states one arrives at, 
\begin{equation}
    \begin{split}
    H_{\mu\nu} = \sum_X   \bigg[ \mel{p(\vb{Q})}{J_\mu}{X(\vb{k})}  & \frac{1}{\omega- (E_X-m_p) +\iu 0} \mel{X(\vb{k})}{ J_\nu}{p(\vb{0})} \\
    &+ 
    \mel{p(\vb{Q})}{J_\nu}{X(-\vb{k}')}  \frac{1}{-\omega'- (E_X-m_p) +\iu 0} \mel{X(-\vb{k}')}{ J_\mu}{p(\vb{0})} \bigg]~, 
    \end{split}
\end{equation}
where $\omega^{(\prime)}= |\vb{k}^{(\prime)}|$ (for on-shell photons) and the states $\ketbra{X}$ are labeled by their center of mass momentum. Let us now consider only a subset of possible intermediate states. Obvious candidates are the elastic channel $\ket{p}$, but other resonance and multi-nucleon channels are also possible. For example any state of the form $\ket{p(\vb{\ell}_1 p(\vb{\ell}_2) \bar{p}(\vb{\ell}_3)}$ where $\vb{\ell}_1+\vb{\ell}_2 +\vb{\ell}_3 = \vb{k}$ or $\vb{Q}-\vb{k}$ is a legitimate intermediate state. 

Including just the $\ket{p}$ and $\ket{pp\bar{p}}$ intermediate states one finds four contributions (two time orderings for both $\ket{p}$ and $\ket{pp\bar{p}}$ intermediate states). The $\ket{p}$ intermediate states involve space-like form factors as one would intuitively expect, but non-covariant propagators i.e., not of the form $(\slashed{p}+m)/(p^2-m^2)$. The $\ket{pp\bar{p}}$ intermediate states involve time-like matrix elements. For example using the first time ordering we encounter the intermediate state $\ket{\vb{p}(\vb*{0})p(\vb{Q}) \bar{p}(\vb{k}-\vb{Q})}$. The time-ordered graph proceeds through pair production from the first current matrix element, $\mel{p(\vb{Q})\bar{p}(\vb{k}-\vb{Q}}{J_\nu}{0}$ followed by annihilation with the second current $\mel{0}{J_\mu}{p(\vb{0})\bar{p}(\vb{k}-\vb{Q})}$. These matrix elements are evaluated using the current $J_\mu= J_\mu(x=0)$, and are defined on-shell. They therefore involve form factors evaluated at the space-like values $q^2\geq 4 m_p^2$ (see Ref.~\cite{Brodsky:1983ta} for a related discussion). 

In direct analogy with the bound-state calculation, if we apply the same method to Compton scattering from an electron, the space-like and time-like form factors are identical (being given just by $F_1=(-e)$ and $F_2=0$). In this case the contributions from the $\ket{e}$ and $\ket{ee\bar{e}}$ states combine and produce the standard covariant expression obtained from covariant perturbation theory. Thus Compton scattering furnishes a simple and analogous example where one can understand the role of time-like and space-like form factors and the contribution of anti-particle states.

\bibliographystyle{apsrev4-1}
\bibliography{biblio}

\end{document}